\documentstyle[aps,prc,twocolumn,floats,epsfig]{revtex}
  \draft
\renewcommand{\vec}[1]{\textbf{#1}}

\newcommand{\rvec}{\vec{r}}
\newcommand{\rvecp}{\vec{r}'}

\newcommand{\sigmavec}{\bbox{\sigma}}

\newcommand{\kvec}{\vec{k}}

\newcommand{\Pop}{\hat{P}}

\newcommand{\nn}{\nonumber}

\newcommand{\MeV}{{\rm MeV}}
\newcommand{\fm}{{\rm fm}}
\newcommand{\tfrac}[2]{{\textstyle\frac{{#1}}{{#2}}}}
\newcommand{\half}{\tfrac{1}{2}}
\newcommand{\iunit}{\text{i}}

\newcommand{\Jt}{{\cal J}^{(2)}}
\newcommand{\etal}{\emph{et al.}}
\newcommand{\ansatz}{\emph{ansatz}}

\newcommand{\rhonu}{\rho_{{\rm n}\uparrow}}
\newcommand{\rhond}{\rho_{{\rm n}\downarrow}}
\newcommand{\rhopu}{\rho_{{\rm p}\uparrow}}
\newcommand{\rhopd}{\rho_{{\rm p}\downarrow}}
\newcommand{\taunu}{\tau_{{\rm n}\uparrow}}
\newcommand{\taund}{\tau_{{\rm n}\downarrow}}
\newcommand{\taupu}{\tau_{{\rm p}\uparrow}}
\newcommand{\taupd}{\tau_{{\rm p}\downarrow}}

\newcommand{\kinfac}{\beta} 

\newcommand{\EF}{{\cal E}}
\newcommand{\HF}{{\cal H}}

\newcommand{\switchJ}{\eta_J}
\begin{document}
\renewcommand{\thefootnote}{\arabic{footnote}}
\twocolumn[\columnwidth\textwidth\csname@twocolumnfalse\endcsname
\title{Gamow-Teller strength and the spin--isospin coupling constants \\ 
      of the Skyrme energy functional
}
\author{M. Bender,$^{1-4}$
         J. Dobaczewski,$^{5,6}$
         J. Engel,$^{3}$
         W. Nazarewicz$^{1,2,5}$
}
\address{$^1$Department of Physics and Astronomy,
              The University of Tennessee,
              Knoxville, Tennessee 37996
}
\address{$^2$Physics Division,
              Oak Ridge National Laboratory,
              P. O. Box 2008, Oak Ridge, Tennessee 37831
}
\address{$^3$Department of Physics and Astronomy,
              The University of North Carolina,
              Chapel Hill, NC 27599
}
\address{$^4$Service de Physique Nucl{\'e}aire Th{\'e}orique,
              Universit{\'e} Libre de Bruxelles,
              CP 229, B-1050 Bruxelles, Belgium
}
\address{$^5$Institute of Theoretical Physics,
              Warsaw University,
              ul. Ho\.za 69, PL-00681, Warsaw, Poland
}
\address{$^6$Joint Institute for Heavy Ion Research,
              Oak Ridge National Laboratory,
              P.O. Box 2008, Oak Ridge, Tennessee 37831
}
\date{December 17, 2001}
\maketitle
\addvspace{5mm}
%
%
\begin{abstract}
We investigate the effects of the spin-isospin channel of the Skyrme
energy functional on predictions for Gamow-Teller distributions and
superdeformed rotational bands.  We use the generalized
Skyrme interaction SkO' to describe even-even ground states and then
analyze the effects of time-odd spin-isospin couplings, first term by
term and then together via linear regression.  Some terms affect the
strength and energy of the Gamow-Teller resonance in finite nuclei
without altering the Landau parameter $g_0'$ that to leading order
determines spin-isospin properties of nuclear matter. Though the existing
data are not sufficient to uniquely determine all the spin-isospin
couplings, we are able to fit them locally.  Altering these coupling
constants does not change the quality with which the Skyrme functional
describes rotational bands.
\end{abstract}
\addvspace{5mm}]
\narrowtext
%
%
\section{Introduction}
Effective interactions for self-consistent nuclear structure
calculations are usually adjusted to reproduce ground-state properties
in even-even nuclei \cite{Ring}. These properties depend only on terms 
in the corresponding energy functional that are bilinear in 
time-reversal-even (or ``time-even'') densities and currents \cite{Eng75a}. 
But the functional also contains an equal number of terms bilinear 
in time-odd densities and currents (see \cite{Eng75a,Dob95c} 
and refs.\ quoted therein), and these terms are seldom independently 
adjusted to experimental data. (For the sake simplicity 
we refer below to terms in the functional as time-even or time-odd, 
even though strictly speaking we mean the densities and currents on 
which they depend.)  The time odd terms can be important as soon as 
time-reversal symmetry (and with it Kramers
degeneracy) is broken in the intrinsic frame of the nucleus. Such
breaking obviously occurs for rotating nuclei, in which the current and
spin-orbit time-odd channels (linked to time-even channels by the gauge
symmetry) play an important role. Time-odd terms also interfere
with pairing correlations in the masses of odd-$A$ and odd-odd
nuclei \cite{Sat98,Xu99a,Rut99a} and contribute to single-particle
energies \cite{Ber80a,Rut98a,Mol00} and magnetic moments \cite{Lip77a}.
Finally, the spin-isospin channel of the effective interaction determines
distributions of the Gamow-Teller (GT) strength.

The latter are the focus of this paper.  We explore the effects of
time-odd couplings on GT resonance energies and strengths, with an eye
toward fixing the spin-isospin part of the Skyrme interaction.
As discussed in our previous study \cite{Eng99a},
there are many good reasons for looking at this channel first.
For instance, a better description of
the GT response should enable more reliable predictions for
$\beta$-decay half-lives of very neutron-rich nuclei.  Those predictions
in turn may help us identify the astrophysical site of r-process
nucleosynthesis, which produces about half of the heavy nuclei
with \mbox{$A>70$}.

Our goal is an improved description of GT excitations in a
fully self-consistent mean-field model. To this end, we treat excited
states in the Quasiparticle Random Phase Approximation (QRPA), with the
residual interaction taken from the second derivative of the energy
functional with respect to the density matrix.
   This approach is equivalent to the small-amplitude limit
of time-dependent Hartree-Fock-Bogoliubov (HFB) theory.  We proceed by
taking the time-odd coupling constants in the Skyrme energy functional
to be free parameters that we can fit to GT distributions.  We then
check that the coupling constants so deduced do not spoil the
description of superdeformed (SD) rotational bands.

Our formulation is nonrelativisitic.  In relativistic mean-field theory
(RMF) \cite{Rin96,Rei89a}, the time-odd
channels, referred to as ``nuclear magnetism,'' are not independent
from the time-even ones because they arise from the small components of
the Dirac wave functions.  For rotational states, the time-odd effects
have been extensively tested and shown to be important for reproducing
experimental data (see, e.g.,\ Ref.\ \cite{Afa00}). Only the current
terms  and spin-orbit terms  play a role there, however, and the time-odd
spin and  spin-isospin channels of the RMF have never been tested
against experimental data.

This paper is structured as follows: In Section \ref{Sect:Skyrme}
we review properties of the Skyrme energy functional. Section
\ref{Sect:Existing} reviews existing parameterizations of
the functional, with particular emphasis on time-odd terms.
Our main results are in Section \ref{Sect:GT}, where we present
calculations of GT strength and discuss the role played by
the time-odd coupling constants. Section \ref{Sect:SD} describes
calculations of moments of inertia for selected SD bands.
Section \ref{Sect:SCO} contains our conclusions. We supplement
our results with six Appendices that provide
more detailed information on local densities and currents
(Appendix \ref{Sect:app:dens}), early parameterizations of
time-odd Skyrme functionals (Appendix \ref{Sect:app:SF}), the limit of
the infinite nuclear matter (Appendix \ref{Sect:app:INM}), Landau
parameters of Skyrme functionals (Appendix \ref{Sect:app:landau})
and of the Gogny force (Appendix \ref{Sect:app:landau:gogny}),
and the residual interaction in finite nuclei from Skyrme functionals
(Appendix \ref{Sect:app:resint}).
%
%
\section{A generalized Skyrme energy functional}
\label{Sect:Skyrme}
%
%
\subsection{Basics of energy density theory}
Many calculations performed with the Skyrme interaction can be viewed as
energy-density theory in the spirit of the Hohenberg-Kohn-Sham
approach \cite{Hoh64a}, originally introduced for many-electron
systems. Nowadays, energy density theory is a standard tool in atomic,
molecular, cluster, and solid-state physics \cite{Nag98a}, as well
as in nuclear physics \cite{Pet91aB}.  The starting point is an energy
functional ${\cal E}$ of all local densities and currents $\rho$,
$\tau$, $\tensor{J}$, $\vec{s}$, $\vec{T}$, and $\vec{j}$ that can
be constructed from the most general single-particle density matrix
\begin{equation}
\label{eq:densitymatrix}
\hat\rho
\equiv \rho (\vec{r}, \sigma, \tau; \vec{r}', \sigma', \tau')
= \sum_k v_k^2 \; \psi_k^* (\vec{r}', \sigma', \tau') \;
                \psi_k (\vec{r}, \sigma, \tau)
\end{equation}
(see Appendix~\ref{Sect:app:dens} for more details), where $\vec{r}$,
$\sigma$, and $t$ are the spatial, spin, and isospin coordinates of the
wave function. The Hohenberg-Kohn-Sham approach maps the nuclear
many-body problem for the ``real'' highly correlated many-body wave
function on a system of independent particles in so-called Kohn-Sham
orbitals $\psi_k$. The equations of motion for $\psi_k$ are derived
from the variational principle
\begin{equation}
\delta {\cal E}
= 0
\quad \Rightarrow \quad
\hat{h} \, \psi_k (\vec{r}, \sigma, \tau)
= \epsilon_k \, \psi_k (\vec{r}, \sigma, \tau)~,
\end{equation}
where the single-particle Hamiltonian $\hat{h}$ is the sum of the
kinetic term $\hat{t}$ and the self-consistent potential $\Gamma$
that is calculated from the density matrix
\begin{equation}
\hat{h}
= \frac{\delta {\cal E}}{\delta \hat\rho}
= \hat{t} + \hat\Gamma [\hat\rho]
\quad .
\end{equation}
The existence theorem for the effective energy functional makes no
statement about its structure. The theoretical challenge is to find an
energy functional that incorporates all relevant physics with as few
free parameters as possible. The density functional approach as used
here is equivalent to the local density approximation to the
nuclear $G$ matrix \cite{Neg72a}.

The energy functional investigated here in detail describes the
particle-hole channel of the effective interaction only.
For the treatment of pairing correlations, the energy functional has
to be complemented by an effective particle-particle interaction
that is constructed in a similar way from the pairing density matrix;
see \cite{Dob84a} for details. We use here the simplest functional
proportional to the square of the local pair density with the
coupling constants given in \cite{Eng99a}.
%
%
\subsection{The Skyrme energy functional}
Within the local-density approximation, the energy functional is given
by the spatial integral of the local energy density ${\cal H}(\vec{r})$
\begin{equation}
{\cal E}
= \int \! d^3 \bbox{r} \; {\cal H}(\vec{r})~.
\end{equation}
The energy density is composed of the kinetic term ${\cal H}_{\rm kin}$,
the Skyrme energy density ${\cal H}_{\rm Skyrme}$ that describes the
effective strong interaction between the nucleons, and a term arising
from the
electromagnetic interaction ${\cal H}_{\rm em}$:
\begin{equation}
{\cal H}
=   {\cal H}_{\rm kin}
   + {\cal H}_{\rm Skyrme}
   + {\cal H}_{\rm em} ~.
\end{equation}
For the electromagnetic interaction, we take the standard Coulomb
expression,
including the Slater approximation for the exchange
term. The energy functional discussed here contains all possible
terms bilinear in local densities and currents and up to second order
in the derivatives that are invariant under reflection,
time-reversal, rotation, translation, and isospin rotation
\cite{Dob96d}.

Time-reversal invariance requires the energy density
to be bilinear in either time-even densities or time-odd densities,
so the Skyrme energy density can be separated into a ``time-even"
part ${\cal H}^{\rm even}$ and a ``time-odd" part ${\cal H}^{\rm odd}$:
\begin{eqnarray}
\label{eq:SkyrmeFu}
{\cal H}_{\rm Skyrme}
& = & \sum_{t = 0,1} \sum_{t_3 = -t}^{t}
       \Big(   {\cal H}_{t t_3}^{\rm even}
             + {\cal H}_{t t_3}^{\rm odd}
       \Big) ~.
\end{eqnarray}
The sum runs over the isospin $t$ and its third component $t_3$.
Only the \mbox{$t_3 = 0$} component of the isovector \mbox{$t = 1$}
terms contribute to nuclear ground states and the rotational bands
discussed later, while the \mbox{$t_3 = \pm 1$} components contribute
only to charge-exchange (e.g.\ GT) excitations. In the notation of
Refs.\  \cite{Dob95c,Dob96d},
the time-even and time-odd Skyrme energy densities read
\begin{eqnarray}
\label{eq:SkyrmeFu:even}
{\cal H}_{t t_3}^{\rm even}\!
& = &\!
         C_t^{\rho} \, \rho_{t t_3}^{2}
       + C_t^{\Delta \rho} \, \rho_{t t_3} \Delta \rho_{t t_3}
       + C_t^{\tau} \, \rho_{t t_3} \tau_{t t_3}
       \nn \\
& + &\!
         C_t^{\nabla J} \rho_{t t_3} \, \nabla \cdot \vec{J}_{t t_3}
       + C_t^{J} \tensor{J}^2_{t t_3},
       \\[2ex]
\label{eq:SkyrmeFu:odd}
{\cal H}_{t t_3}^{\rm odd}\!
& = &\!
         C_t^{s} \, \vec{s}^{2}_{t t_3}
       + C_t^{\Delta s} \, \vec{s}_{t t_3} \cdot \Delta \vec{s}_{t t_3}
       + C_t^{T} \, \vec{s}_{t t_3} \cdot \vec{T}_{t t_3}
       \nn \\
& + &\!
         C_t^{\nabla s} \, (\nabla \cdot \vec{s}_{t t_3})^2
       + C_t^{j} \, \vec{j}^2_{t t_3}
       + C_t^{\nabla j} \,
         \vec{s}_{t t_3} \cdot \nabla \times \vec{j}_{t t_3}.\!\!
\end{eqnarray}
Isospin invariance of the Skyrme interaction makes the coupling
constants independent of the isospin $z$-projection.
All coupling constants might be density dependent. Following the
standard \ansatz\ for the Skyrme interaction, we neglect such a
possibility
except in $C_t^{\rho}$ and $C_t^{s}$, for which we restrict the density
dependence to the following form
\begin{eqnarray}
\label{eq:ddeven}
C_t^{\rho} [\rho_0]
& = & C_1^{\rho}[0]
       + \Big(C_1^{\rho}[\rho_{\rm nm}]-C_1^{\rho}[0]\Big)
       \left( \frac{\rho_0}{\rho_{\rm nm}} \right)^{\alpha}~,
       \\
\label{eq:ddodd}
C_t^{s} [\rho_0]
& = &  C_1^{s}[0]
       + \Big(C_1^{s}[\rho_{\rm nm}]-C_1^{s}[0]\Big)
       \left( \frac{\rho_0}{\rho_{\rm nm}} \right)^{\xi} .
\end{eqnarray}
Here $\rho_0$ is the isoscalar scalar density, and $\rho_{\rm nm}$ is its
value in saturated infinite nuclear matter. The exponent  $\alpha$ that
specifies the density dependence of $C_t^{\rho} [\rho_0]$ must be
about 0.25 for the incompressibility coefficient $K_\infty$ to be
correct\cite{Fri86a,Bla95a,Cha97a,Cha98a}. Although this fact does
not restrict the analogous power in $C_t^{s} [\rho_0]$, Eq.\
(\ref{eq:ddodd}), we keep $\xi$ equal to $\alpha$ for simplicity here.
Usually we will consider energy functionals that
are invariant under local gauge transformations \cite{Dob95c}, which
generalize the Galilean invariance of the Skyrme interaction discussed
in
\cite{Eng75a}. Gauge invariance links three pairs of time-even and
time-odd terms in the energy functional:
\begin{eqnarray}
\label{eq:gauge}
C_t^{j}        & = & - C_t^{\tau}     \quad , \quad
C_t^{J}          =   - C_t^{T}        \quad , \quad
C_t^{\nabla j}   =   + C_t^{\nabla J} ~,
\end{eqnarray}
These relations fix all orbital time-odd terms, leaving only
time-odd terms corresponding to the spin-spin interaction with free
coupling constants. Relations (\ref{eq:gauge}) lead to a simplified
form of Eqs.\ (\ref{eq:SkyrmeFu})--(\ref{eq:SkyrmeFu:odd}),
\begin{eqnarray}
\label{eq:SkyrmeFu:gauge}
{\cal H}_{\rm Skyrme}
& = &  \sum_{t = 0,1} \sum_{t_3 = -t}^{t} \Big[
         C_t^{\rho} \, \rho_{t t_3}^{2}
       + C_t^{s} \, \vec{s}^{2}_{t t_3}
       \nn \\
& + &
         C_t^{\Delta \rho} \, \rho_{t t_3} \Delta \rho_{t t_3}
       + C_t^{\Delta s} \, \vec{s}_{t t_3} \cdot \Delta \vec{s}_{t t_3}
       \nn \\
& + &
         C_t^{\tau} \, ( \rho_{t t_3} \tau_{t t_3} - \vec{j}^2_{t t_3} )
       + C_t^{T} \,
         (   \vec{s}_{t t_3} \cdot \vec{T}_{t t_3}
           - \tensor{J}^2_{t t_3} )
       \nn \\
& + &
         C_t^{\nabla J} \,
         (   \rho_{t t_3} \, \nabla \cdot \vec{J}_{t t_3}
           + \vec{s}_{t t_3} \cdot \nabla \times \vec{j}_{t t_3} )
       \nn \\
& + &
         C_t^{\nabla s} \, (\nabla \cdot \vec{s}_{t t_3})^2 \Big]
         \quad .
\end{eqnarray}
The time-even terms of the energy functional can be directly
related to nuclear bulk properties such as $E/A$, the saturation density
$\rho_{\rm n.m.}$, incompressibility, symmetry energy, surface and
surface symmetry energy, and spin-orbit splittings.  The remaining
time-odd terms cannot.

We will set the coupling constant
\mbox{$C_t^{\nabla s}$} to 0.  The term it multiplies comes
from a local two-body tensor force considered in Skyrme's original
papers \cite{Sky56a} and discussed by Stancu \etal\ \cite{Sta77},
but omitted in all modern Skyrme parameterizations except the force SL1
introduced by Liu \etal\ \cite{Liu91a}, which has not been used since.
%
%
\section{Existing parameterizations}
\label{Sect:Existing}
The coupling constants of the time-odd Skyrme energy
functional are usually taken from the (antisymmetrized)
expectation value of a Skyrme force \cite{Eng75a}. When so obtained, the
16 coupling constants of the energy functional
(\ref{eq:SkyrmeFu:gauge}) are uniquely linked to the 10 parameters
$t_i$, $x_i$, $W_0$, and $\alpha$ of the standard Skyrme force (see
Appendix~\ref{Sect:app:SF} and Eq.\ (\ref{eq:cpl:SF})).
Only a few parameterizations rigidly enforce these relations,
however. Among them are
the forces of Ref.\ \cite{Fri86a} (e.g.\ Z$_\sigma$), SkP \cite{Dob84a},
the Skyrme forces of Tondeur \cite{Ton84a}, the recent parameterizations
SLy5 and SLy7 \cite{Cha98a}, and SkX \cite{Bro98a}.  Most other
parameterizations neglect the $\tensor{J}^2$ term obtained from the
two-body Skyrme force, setting \mbox{$C_t^T = 0$}.  Some authors do this
for practical reasons; the $\tensor{J}^2$ term is time-consuming to
calculate, and its contribution to the total binding energy is rather
small.  Other authors (see, e.g., \cite{Bei75a}) find that including it
with a coupling dictated from the HF expectation value of the Skyrme
force can lead to unphysical solutions and/or unreasonable spin-orbit
splittings.  For spherical shapes, the $\tensor{J}^2$ term contributes
to the time-even energy density in the same way as the neglected tensor
force.  One might therefore argue that by including the tensor force one
could counterbalance the unwanted $\tensor{J}^2$ term exactly
\cite{Bei75a}.  This argument, however, applies neither to deformed
shapes nor to time-odd fields. Moreover, neglecting this term
often violates self-consistency on the QRPA level (see below).

Although one might disagree with the rationale for neglecting the
$\tensor{J}^2$ terms, it is not easy to adjust the coupling
constants $C_t^T$ to spectral data.  Large values for
$C_t^T$ can be ruled out because they spoil the previously
obtained agreement for single-particle spectra, but there are
broad regions of values where they influence the usual time-even
observables too weakly to be uniquely determined \cite{PGRpc}.
Only once in the published literature has there been an attempt
to do so \cite{Ton83a}.

All first-generation Skyrme interactions, e.g.,  SI, SII \cite{Vau72a},
and SIII \cite{Bei75a}, used a three-body delta force instead of a
density-dependent two-body delta-force to obtain reasonable
nuclear-matter properties. The three-body interactions led
to \mbox{$\alpha = 1$} for $C_t^\rho$ in Eq.\ (\ref{eq:ddeven}),
but a different density dependence of the $C_t^s$.
\mbox{$\alpha = 1$} is too large to get the
incompressibility $K_\infty$ right, and causes a spin instability
in infinite nuclear matter \cite{Cha75a} and finite nuclei
\cite{Str76a} (again only within a microscopic potential framework).
Both problems are cured with smaller values of $\alpha$ (between $1/6$
and $1/3$ \cite{Cha97a}) but the second-generation interactions that did
so still had problems in the time-odd channels, giving a poor
description of spin and spin-isospin excitations and prompting several
attempts to describe finite nuclei with extended Skyrme interactions.
Krewald \etal\ \cite{Kre77a}, Waroquier \etal\ \cite{War83a}, and Liu
\etal\ \cite{Liu91a}, for example, introduced additional three-body
momentum-dependent forces.  Waroquier \etal\ added an admixture of the
density-dependent two-body delta force and a three-body delta force,
while Liu \etal\ considered a tensor force. But none of these
interactions has been used subsequently.

Van Giai and Sagawa \cite{Gia81a} developed the more durable
parameterization SGII, which gave a reasonable description of GT
resonance data known at the time and is still used today.  The fit to
ground state properties was made without the $\tensor{J}^2$ terms,
however, even though they were used in the QRPA. Consequently,
in such an approach, the QRPA does not correspond to the
small-amplitude limit of time-dependent HFB.

All these attempts to improve the description of the
time-odd channels impose severe
restrictions on the coupling by linking them to the HF expectation value
of a Skyrme force, leading to one difficulty or another.  The authors of
Refs.\ \cite{Neg72a,Rei85a} proceed differently,
treating the Skyrme energy functional as the result of a local-density
approximation. The interpretation of the Skyrme interaction as
an energy-density functional, besides relaxing the restrictions on the
time-odd couplings, endows the spin-orbit interaction with
a more flexible isospin structure \cite{Lal94a,Sha95a,Rei95a}
than can be obtained from the standard Skyrme force \cite{Bel56a}.
Some of the parameterizations used here will take advantage of that
freedom.  But the authors of Ref.\ \cite{Rei85a} include only
time-odd terms that are determined  by gauge invariance; the other
couplings are tentatively set to zero
(\mbox{$C_t^s = C_t^{\Delta s} = 0$}).
Such a  procedure is reasonable when describing natural parity
excitations within the (Q)RPA, but the neglected spin-spin terms are
crucial for the unnatural parity states that we discuss.

In this study,  we use the energy-functional approach
(\ref{eq:SkyrmeFu:gauge}) with fully independent time-even and time-odd
coupling constants. Our hope is that this more general formulation will
improve the description of the GT properties while leaving the good
description of ground-state properties in even nuclei untouched.
%
%
\section{Giant Gamow--Teller resonances}
\label{Sect:GT}

The repulsive interaction between proton particles and neutron holes
in the \mbox{$J^{\pi}=1^+$} (spin-isospin) channel gives rise to a giant
charge-exchange resonance in all nuclei with excess neutrons.  The
centroid of the resonance (which typically has a width
of 5-10 MeV) can be roughly parameterized by the simple formula
\mbox{$E_{\rm GT} - E_{\rm F} = 26 A^{-1/3} - 18.5 (N-Z) A^{-1}$},
where $E_{\rm F}$ is the centroid of the Fermi resonance \cite{Nak82a}.
This formula, however, captures only average behavior; individual cases
depend on single-particle structure,
and in particular the spin-orbit splitting.

The ability to model GT resonances is crucial for predictions
of nuclear $\beta$ decay.  Just as the low-lying $E1$ strength is
depleted by the giant dipole resonance, so the low-lying GT
strength, responsible for $\beta$ decay, is affected by the GT
resonance. Since one of our future goals is an improved calculation
of $\beta$-decay rates in nuclei along the $r$-process path, it is
important to develop a reliable description of the GT giant
resonance.
%
%
\subsection{Residual interaction in finite nuclei}
Non-self-consistent calculations often use the residual
Lan\-dau-Mig\-dal inter\-action in the spin-isospin channel:
\begin{eqnarray}
\label{eq:resint:landau}
\lefteqn{
v_{\rm res} (\vec{r}, \vec{r}')
} \nn \\
& = & N_0
       \Big[   g_0' \, \delta (\vec{r} - \vec{r}')
               + g_1' \, \vec{k}' \cdot \delta (\vec{r} - \vec{r}') \,
\vec{k}
       \Big]
       (\bbox{\sigma} \cdot \bbox{\sigma}') \,
       (\bbox{\tau} \cdot \bbox{\tau}')
\end{eqnarray}
where $N_0$ is a normalization factor [see Eq.\ (\ref{eq:N0})] and
$\vec{k}$ and $\vec{k}'$ are defined in Appendix \ref{Sect:app:SF}.
In most applications, only the $s$-wave interaction with strength $g_0'$
is used, and the matrix elements of the force are not antisymmetrized.
The underlying single-particle spectra are usually taken from
a parameterized potential, e.g., the Woods-Saxon potential.
Typical values for $g_0'$, obtained from fits to GT-resonance
systematics, are \mbox{$1.4 \leq g_0' \leq 1.6 $}
\cite{Ber81a,Gaa81aE,Suz82a}.  (See Ref.\ \cite{Ber81b} for an early
compilation of data.) Sometimes this approach is formulated in terms
of the residual interaction between antisymmetrized states. The results
are similar, e.g.,\ \mbox{$g_0' = 1.54$} in the double-$\beta$-decay
calculations by Engel \etal\ \cite{Eng88a}. More complicated residual
interactions, like boson-exchange potentials, have been used as well;
see, e.g., Refs.~\cite{Tow87a,Ost91a,Ost92a}. Borzov et al. use
a renormalized one-pion exchange potential in connection with a 
\mbox{$\ell = 0$} Landau-Migdal interaction of type 
(\ref{eq:resint:landau}) \cite{Borzov}.

A much simpler residual interaction in the GT channel is
a separable (or ``schematic") interaction, $v_{\rm res} =
\kappa_{\rm GT} \; (\bbox{\sigma} \cdot \bbox{\sigma}' )\;
  (\bbox{\tau} \cdot \bbox{\tau}')$, where the strength $\kappa_{\rm GT}$
has to be a function of $A$. This interaction
is widely used in global calculations of nuclear $\beta$-decay
\cite{Mol90a,Homma}. Sarriguren \etal\ \cite{Sar98a}
use it for a description of the GT resonances in deformed
nuclei with quasiparticle energies obtained from self-consistent
HF+BCS calculations. They estimate $\kappa_{\rm GT}$ from the Landau
parameters of their Skyrme interaction.  (The same prescription is used
in their calculations of $M1$ resonances \cite{Sar96a}.)
But however useful this approach may be from a technical point of
view, it is not self-consistent. Nor is it equivalent to using the
original residual Skyrme interaction; see, e.g.,\ the
discussion in \cite{Gaa81aE}.

A truly self-consistent calculation, by contrast, should interpret
the QRPA as the small-amplitude limit of time-dependent HFB theory.
The Skyrme energy functional used in the HFB should then determine
the residual interaction between unsymmetrized states in the QRPA:
\begin{equation}
\label{eq:d2Edrhodrho}
v_{\rm res}
= \frac{\delta^2 {\cal E}}
        {\delta \rho (\vec{r}_1, \sigma_1, \tau_1; \vec{r}_2, 
\sigma_2, \tau_2) \,
         \delta \rho (\vec{r}_1', \sigma_1', \tau_1'; \vec{r}_2', \sigma_2',
\tau_2')} ~.
\end{equation}
The actual form of the residual interaction that contributes to the
QRPA matrix elements of $1^+$ states is outlined in Appendix
\ref{Sect:app:resint}.
%
%
\begin{table}[t!]
\caption{\label{tab:landau}
Landau parameters for various Skyrme interactions from relations
(\protect\ref{eq:cpl:SF}) and the Gogny forces D1 and D1s.
Missing entries are zero by construction.
}
\begin{tabular}{ldddddd}
Force & $g_0$ & $g_1$ & $g_2$ &$g_0'$ & $g_1'$ & $g_2'$ \\
\noalign{\smallskip}\hline\noalign{\smallskip}
SkM*  &  0.33 &       &      & 0.94 &      &       \\
SGII  &  0.62 &       &      & 0.93 &      &       \\
SkP   & -0.23 & -0.18 &      & 0.06 & 0.97 &       \\
SkI3  &  1.89 &       &      & 0.85 &      &       \\
SkI4  &  1.77 &       &      & 0.88 &      &       \\
SLy4  &  1.39 &       &      & 0.90 &      &       \\
SLy5  &  1.14 &  0.24 &      &-0.15 & 1.05 &       \\
SLy6  &  1.41 &       &      & 0.90 &      &       \\
SLy7  &  0.94 &  0.47 &      & 0.02 & 0.88 &       \\
SkO   &  0.48 &       &      & 0.98 &      &       \\
SkO'  & -1.61 &  2.16 &      & 0.79 & 0.19 &       \\
SkX   & -0.63 &  0.18 &      & 0.51 & 0.53 &       \\
\noalign{\smallskip}\hline\noalign{\smallskip}
D1  & 0.47 &  0.06 & 0.12 & 0.60 & 0.34 &  0.08 \\
D1s & 0.48 & -0.19 & 0.25 & 0.62 & 0.62 & -0.04
\end{tabular}
\end{table}
%
%
%
\subsection{GT strength distributions from existing Skyrme interactions}
\label{Subsect:GT:orig}
Before exploring the time-odd degrees of freedom of the generalized
Skyrme energy functional, we analyze the performance of
existing parameterizations when relations (\ref{eq:cpl:SF}) are used.
We examine the forces SkP \cite{Dob84a}, SGII \cite{Gia81a},
SLy4, SLy5 \cite{Cha98a}, SkO, and SkO'\cite{SkO}, which all provide a
good description of ground-state properties but differ in details.
SkP uses an effective mass \mbox{$m^*/m = 1$} and is designed to
describe both the mean-field and pairing effects%
\footnote{Since the effective mass scales the average density of
single-particle states, it might visibly influence the GT
strength distribution.
}.
All other forces have smaller effective masses, so that
\mbox{$m^*/m \approx 0.9$} (SkO$x$) or even \mbox{$m^*/m \approx 0.7$}
(SGII, SLy$x$).   SGII represents an early attempt to get good
GT response properties from a standard Skyrme force. SLy4 and SLy5 are
attempts to reproduce properties of pure neutron matter together with
those of normal nuclear ground states.  SkO and SkO' are recent fits that
include data from exotic nuclei, with particular emphasis on isovector
trends in neutron-rich Pb isotopes; they complement the
spin-orbit interaction with an explicit isovector degree-of-freedom
\cite{Rei95a}. All other parameterizations use the standard prescription
\mbox{$C_0^{\nabla J} = 3 C_1^{\nabla J}$}.

Residual interactions are often summarized by
the Landau parameters that appear in Eq.\ (\ref{eq:resint:landau}).
The parameters can be derived as the corresponding coupling constants
when Eq.\ (\ref{eq:d2Edrhodrho}) is evaluated for infinite spin-saturated
symmetric nuclear matter (see Appendix \ref{Sect:app:landau}).
In the literature, the infinite nuclear matter (INM)
properties of the Skyrme interactions are
usually calculated from Eqs.\ (\ref{eq:cpl:SF}). For the generalized
energy functional (\ref{eq:SkyrmeFu:gauge}) discussed here, the
time-even INM properties such as the saturation density, energy per
particle, effective mass, incompressibility, symmetry coefficient,
and the time-even Landau parameters $f_i$, $f_i'$
are unchanged, but properties of polarized INM and expressions for the
time-odd Landau parameters $g_i$ and $g_i'$ are different.
We derive them in Appendix~\ref{Sect:app:landau}. Here we are most
concerned with the Landau parameters in the spin and spin-isospin
channels,
\begin{mathletters}
\begin{eqnarray}
\label{eq:g0}
g_0
& = & N_0 \big(   2 C_0^{s}
                 + 2 C_0^{T} \, \kinfac \, \rho_0^{2/3}
           \big),
        \\
\label{eq:g0p}
g_0'
& = & N_0 \big(   2 C_1^{s}
                 + 2 C_1^{T} \, \kinfac \, \rho_0^{2/3}
           \big),
        \\
\label{eq:g1}
g_1
& = & - 2 N_0 \; C_0^{T} \, \kinfac \, \rho_0^{2/3},
        \\
\label{eq:g1p}
g_1'
& = & - 2 N_0 \; C_1^{T} \, \kinfac \, \rho_0^{2/3},
\end{eqnarray}
\end{mathletters}
\begin{figure}[t!]
\centerline{\epsfig{file=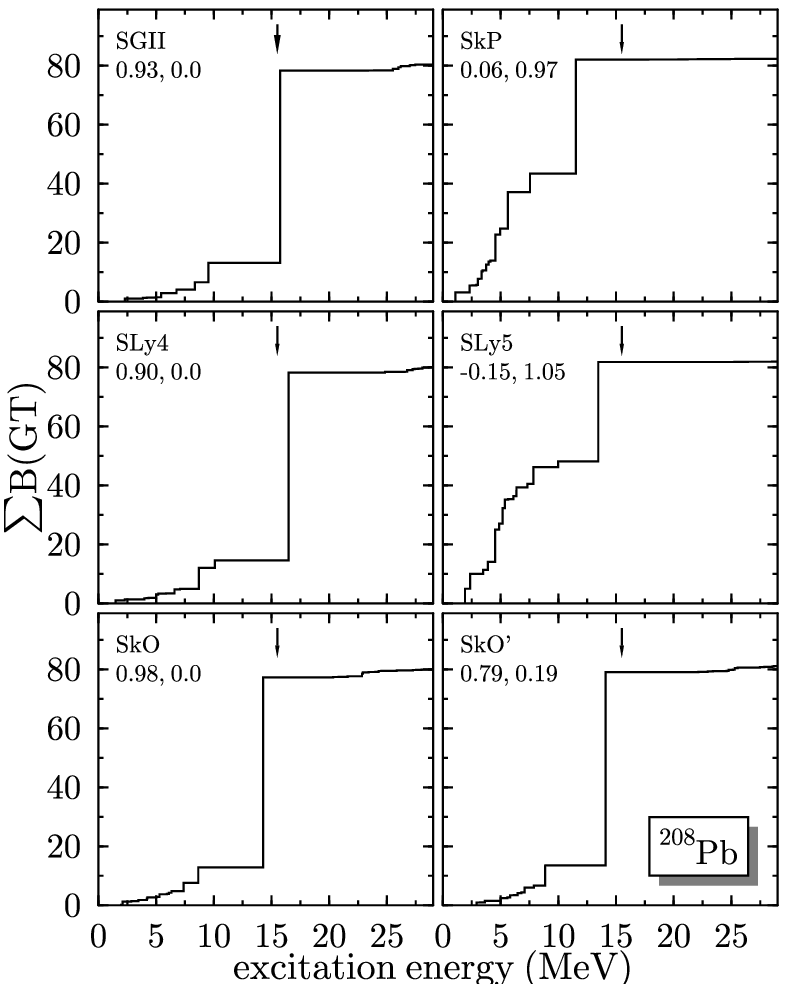}}
\caption{\label{Fig:GT:pb208_mfvergl}
Summed GT strength in $^{208}$Pb calculated with several Skyrme
interactions,
each corresponding to the Landau parameters $g_0'$ and $g_1'$
as indicated. The experimental resonance energy, taken from
Ref.\ \protect\cite{Gaa81aE}, is indicated by an arrow.
}
%
\centerline{\epsfig{file=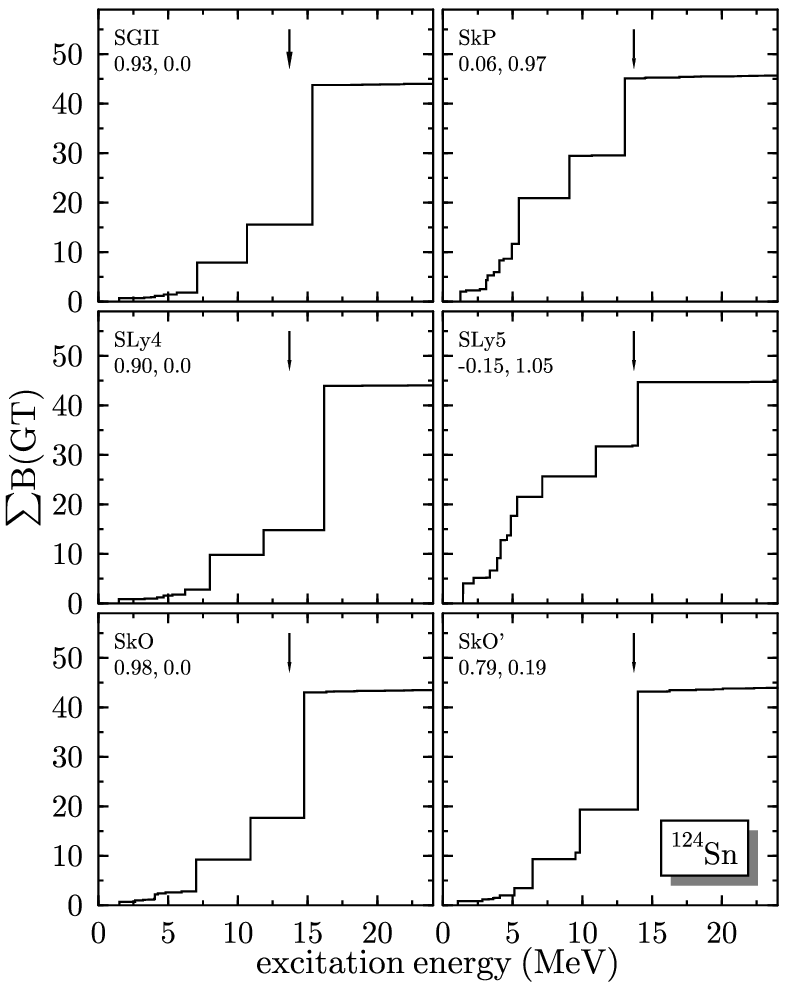}}
\caption{\label{Fig:GT:sn124_mfvergl}
Same as in Fig.~\protect\ref{Fig:GT:pb208_mfvergl}
except for $^{124}$Sn.
}
\end{figure}
where $N_0$ is given by (\ref{eq:N0}) and
$\kinfac = ( 3 \pi^2 / 2 )^{2/3}$. Values for some typical Skyrme
interactions appear in Table \ref{tab:landau}. Higher-order Landau
parameters are zero for the Skyrme functional (\ref{eq:SkyrmeFu:gauge}).
Some of these values differ from those given elsewhere because,
unlike other authors, we insist on exactly the same effective interaction
in the HFB and QRPA.
The coupling constants $C^T_0$ and $C^T_1$ are fixed by the gauge
invariance of the energy functional, which means that \mbox{$C^T_1 = 0$}
for SGII, SLy4 and SkO, because the $\tensor{J}^2$ term was omitted in
the corresponding mean-field fits. For these interactions \mbox{$g_1' = 0$}
and \mbox{$g_0' \approx 0.9$}.
For SkP and SLy5, and SLy7, $C_1^T$ is relatively large (see
Table~\ref{tab:toddcplg}),  leading to a large \mbox{$g_1' \approx 1.0$},
but a cancellation between  two terms makes \mbox{$g_0' \approx 0.0$}.
\begin{figure}[t!]
\centerline{\epsfig{file=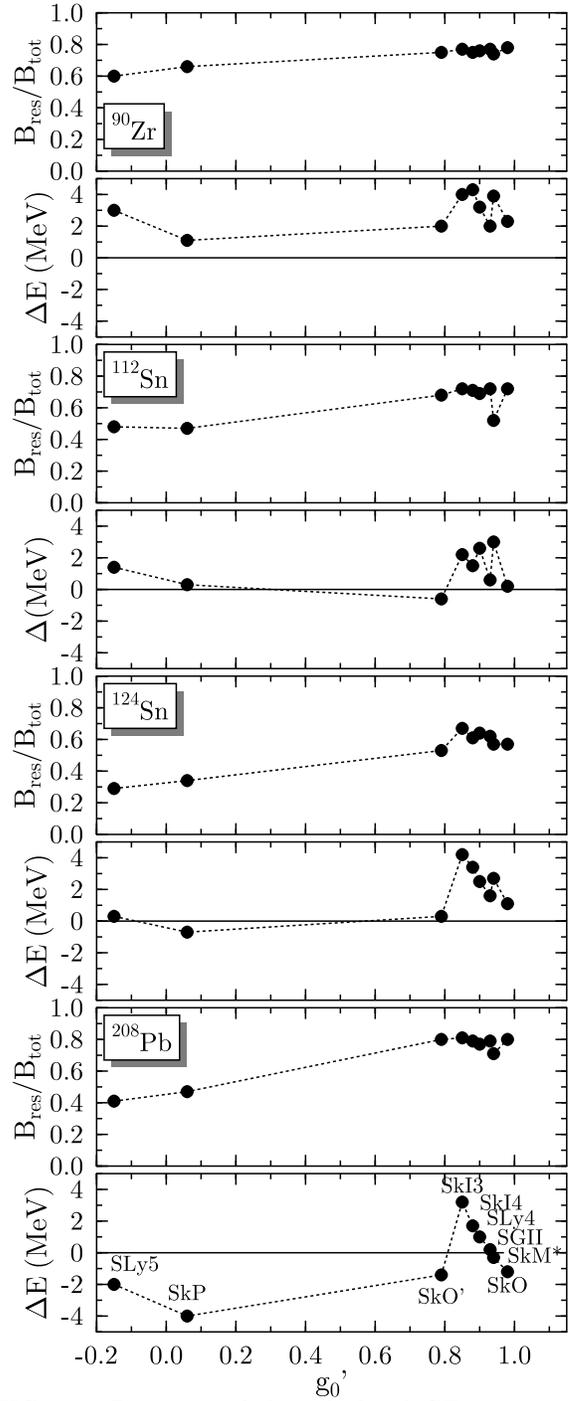}}
\caption{\label{Fig:GT:org}
Deviation of the calculated GT resonance energy from experiment,
$E_{\rm calc} - E_{\rm expt}$, and fraction of the GT strength in
the resonance, $B_{\rm res}/B_{\rm tot}$, versus Landau
parameter $g_0'$, calculated for several Skyrme interactions (as
indicated in the lower right panel) in $^{90}$Zr, $^{112}$Sn,
$^{124}$Sn, and $^{208}$Pb. Experimental values are taken from
Ref.\ \protect\cite{Gaa81aE}.
}
\end{figure}

Table \ref{tab:landau} also gives values for the Landau parameters
calculated for the Gogny forces D1 \cite{Dec80a} and D1s \cite{Bla95a}
from the expressions provided in Appendix \ref{Sect:app:landau:gogny}.
In the spirit of the Gogny force as a two-body potential, one has no
freedom to choose the time-odd terms independently from the time-even
ones. (Note that the Gogny force, however, employs the same local-density
approximation for the density-dependence as the Skyrme energy
functional that contributes to the {$\ell =0$} Landau parameters.)
The higher-order Landau parameters are uniquely fixed by the finite-range
part of the Gogny force.

Figures~\ref{Fig:GT:pb208_mfvergl} and \ref{Fig:GT:sn124_mfvergl} show the
summed GT strength $B(GT)$ in $^{208}$Pb and $^{124}$Sn, calculated with
all the selected Skyrme forces. The ground-state energies are calculated 
as described
in Ref.\ \cite{Eng99a}, and all strengths are divided by $1.26^2$,
following common practice, to account for GT quenching.  Although the GT
resonance in $^{208}$Pb comes out at about the right energy for SGII,
SLy4, SkO, and SkO', it is too low for SkP and SLy5.  These latter two
interactions also leave too much GT strength at small excitation
energies.  It is tempting to interpret these findings in terms of the
Landau parameters for these interactions.  Schematic models suggest
\cite{Ring} that an increase of $g_0'$ results in an increased resonance
energy and more GT strength in the resonance.  The nucleus $^{208}$Pb
indeed behaves in this way, as can be seen
in Fig.~\ref{Fig:GT:pb208_mfvergl}.  The forces SkP and SLy5, with small
values of $g_0'$, yield more low-lying strength and a lower resonance
energy than the remaining forces which correspond to \mbox{$g_0' \approx
0.9$}.
%
%
\begin{figure}[t!]
\centerline{\epsfig{file=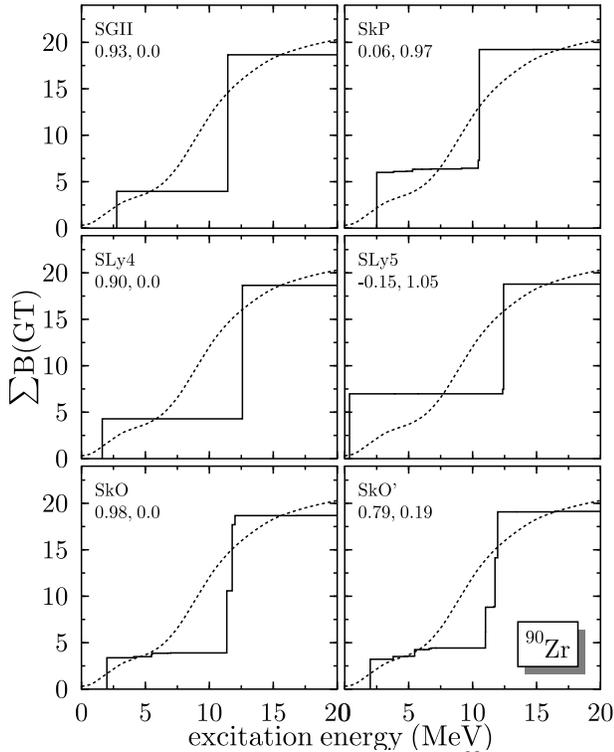}}
\caption{\label{Fig:GT:zr90_mfvergl}
Same as in Fig.~\protect\ref{Fig:GT:pb208_mfvergl} except for $^{90}$Zr.
The very detailed experimental data are from a recent
experiment by Wakasa \protect\etal\ \protect\cite{Wak97aE}.
}
\end{figure}
%
%
\begin{figure}[b!]
\centerline{\epsfig{file=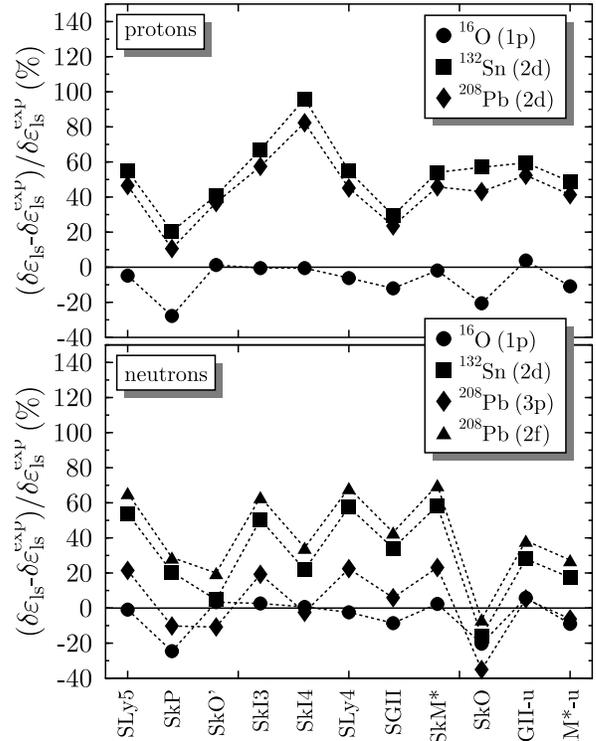}}
\caption{\label{Fig:lssplit}
Relative errors in the spin-orbit splitting (calculated from the
intrinsic single-particle energies) for the forces, nuclei, and states
indicated. Only splittings between states which are both above or
both below the Fermi surface are included. Other states are affected
by core polarization and cannot be safely described by the mean field
\protect\cite{Ber80a,Rut98a}.
The forces SLy5--SkO are ordered according to their values for
$g_0'$ (see Fig.~\protect\ref{Fig:GT:org}). SGII--u and SkM$^*$--u
are two recent forces with modified spin-orbit interactions tailored
for future use in GT resonance studies \protect\cite{Sag01a}.
}
\end{figure}
%
%

In $^{124}$Sn, however, this simple picture does not hold, as
Fig.~\ref{Fig:GT:sn124_mfvergl} shows.  The resonance energies are
similar (and close to the experimental value) for SkP, SLy5, SkO, and
SkO' forces with very different values of $g_0'$, while SGII and SLy4
push the resonance energy too high.  Only the amount of the low-lying
strength seems to scale with $g_0'$.  It is interesting, though, that
the related forces SLy4 and SLy5 (which predict very similar
single-particle spectra, but have quite different GT residual
interactions) agree with the schematic model in that SLy4, with larger
$g_0'$, puts the GT resonance at a higher excitation energy.

It is clear that the scaling predicted by the schematic model is too
simple, and Fig.~\ref{Fig:GT:org} demonstrates this clearly.  There we
show the calculated strengths $B_{\rm res}$ in the GT resonances
relative to the sum-rule value \mbox{$B_{\rm tot}=3(N-Z)$}, and the
calculated GT resonance energies $E_{\rm calc}$ relative to the
experimental values $E_{\rm expt}$.  [For $^{90}$Zr, $^{112}$Sn,
$^{124}$Sn, and $^{208}$Pb we used \mbox{$E_{\rm expt} = 9.4$\,MeV},
8.9\,MeV, 13.7\,MeV, and 15.5\,MeV, respectively \cite{Gaa81aE}.  Note
that the calculated resonance energy depends on a prescription (see
\protect\cite{Eng99a}) not strictly dictated by the QRPA.]  The scatter
near \mbox{$g_0' \approx 0.9$}, in both the resonance energy and in the
amount of low-lying strength, shows that other combinations of parameters
in the residual interaction besides $g_0'$ affect the GT distribution.
This is not entirely surprising given the complexity of finite nuclei
and of the interaction (\ref{eq:d2Edrhodrho}).  In
Sect.~\ref{Subsect:GT:gen} we quantify these other important
combinations and discuss their effects.

But another factor, this one determined by the time-even part of the
Skyrme functional, affects the GT distribution:  the underlying
single-particle spectrum.  Since GT transitions are especially sensitive
to proton spin-orbit splittings, small changes in the time-even part of
the force can, in principle, move the GT resonance considerably.
Sensitivity to the spin-orbit splitting is particularly obvious in
$^{90}$Zr, where detailed information has been obtained from a recent
experiment by Wakasa \etal\ \cite{Wak97aE}.  Unlike in $^{124}$Sn and
$^{208}$Pb, which respond to a GT excitation in a collective way, the
$^{90}$Zr GT spectrum is dominated by two single-particle transitions,
from the neutron $1g_{9/2}$ state to the proton $1g_{9/2}$ and
$1g_{7/2}$ states.  The difference between the locations of the two
peaks in the GT spectrum is the sum of the proton $1g$ spin-orbit
splitting and a contribution from the residual interaction (which can be
expected to increase the difference).  As Fig.~\ref{Fig:GT:zr90_mfvergl}
shows, all interactions, whatever their value for $g_0'$, overestimate
this difference; the resonance energy is always too large, even when the
residual interaction is switched off completely.

Most Skyrme interactions give spin-orbit splittings in heavy nuclei that
are too large \cite{Ben99a}.  We can therefore expect errors in their
predicted GT strength distributions \cite{Ber81a,Suz82a}.
Figure~\ref{Fig:lssplit} shows errors in the predicted spin-orbit energies
for the same forces as in Fig.~\ref{Fig:GT:org}.  Interactions such as
SkI3, SkI4, or SLy4 that overestimate the proton spin-orbit splittings
give the largest resonance energies (and tend to overestimate them).
The best interaction, in view of the combined information from Figs.\
\ref{Fig:GT:org} and \ref{Fig:lssplit}, appears to be SkO'.  Therefore,
below, we use its time-even energy functional for
further exploration of the time-odd terms.

We have included some new forces in Fig.\ \ref{Fig:lssplit}; in a recent
paper \cite{Sag01a}, Sagawa \etal\ attempt to improve the spin-orbit 
interaction for the standard Skyrme forces SIII, SkM$^*$, and SGII, aiming
at better GT-response predictions. They generalize the spin-orbit
interaction through the condition \mbox{$C_0^{\nabla J} = - C_1^{\nabla
J}$} and include the $\tensor{J}^2$ term with a coupling given by
Eq.~(\ref{eq:cpl:SF}). Although the modified forces SkM$^*$-u, and SGII-u
give slightly better descriptions of GT resonances than the original
interactions, they generate unacceptable errors in total binding
energies and do not substantially improve the overall description of
single-particle spectra in $^{208}$Pb.

A few remarks are in order before proceeding:  (i) The spin-orbit
splittings shown in Fig.~\ref{Fig:lssplit} are calculated from intrinsic
single-particle energies.  Since experimental data are obtained from
binding-energy differences between even-even and adjacent odd-mass
nuclei, core polarization induced by the unpaired nucleon, which depends
partly on time-odd channels of the interaction \cite{Ber80a,Rut98a},
alters single-particle energies.  The effect is largest in small nuclei
(of the order of $20 \%$ in $^{16}$O), decreasing rapidly with mass
number \cite{Rut98a}.  (ii) GT distributions are also affected by the
particle-particle channel of the effective interaction, but mainly at
low energies.  The GT resonance is not materially altered \cite{Eng99a},
so we can safely neglect the particle-particle interaction here.
%
%
\subsection{GT resonances from generalized Skyrme functionals}
\label{Subsect:GT:gen}
%

%
\begin{figure}[t!]
\epsfig{file=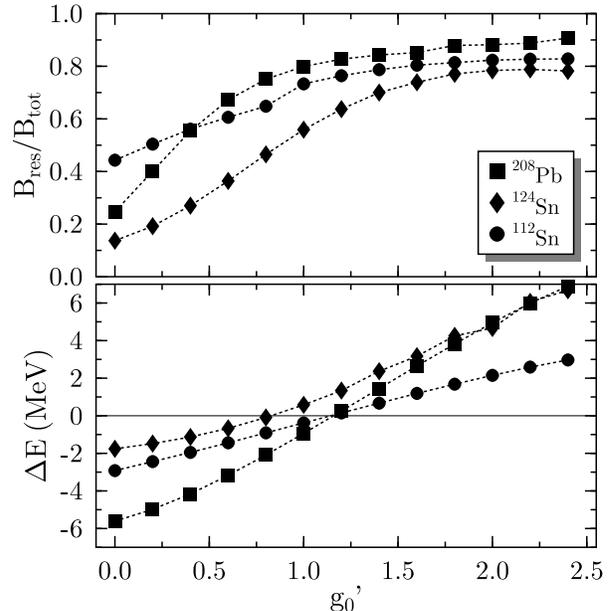}
\caption{\label{fig:sko4p_c0p}
Deviation of calculated and experimental GT resonance energies
(lower panel) and a fraction of the GT strength in the GT resonance
(upper panel) for $^{112}$Sn, $^{124}$Sn, and $^{208}$Pb,
calculated with SkO' and a modified residual
spin-isospin interaction. $C_1^{T}$ is kept at the Skyrme-force
value and $C_1^{\Delta s}$ is set to zero. $C_1^{s}$ is chosen to
be density-independent and varied to get $0 \leq g_0' \leq 2.4$.
\mbox{$g_1' = 0.19$} in all cases.
}
\end{figure}
%
%
We turn now to generalized energy functionals in which the time-odd
coupling constants $C_t^s$, $C_t^{\Delta s}$, and $C_t^T$ are treated
as free parameters; that is, we no longer insist that the interaction
correspond to a two-body Hamiltonian with matrix elements that should
be antisymmetrized. As we showed in Sect.\ \ref{Subsect:GT:orig},
values of the Landau parameter $g_0'$ alone are insufficient to link
the properties of the GT resonance to the coupling constants of the
energy density functional. In this section, using  the time-even
functional of SkO', we study the dependence of the GT
resonance on several other
combinations of the coupling constants as well. Because the isoscalar
time-odd terms do not affect the GT transitions, we focus here on the
isovector coupling constants $C_1^s$, $C_1^{\Delta s}$, and $C_1^T$.
%
%
\subsubsection{Study of $C_1^{s} [\rho_{\rm nm}]$.}
We begin with the simplest case, assuming that
(i) the functional is gauge invariant,
(ii) all time-odd coupling constants are density-independent, and
(iii) the spin-surface term can be neglected, i.e.,
\mbox{$C_1^{\Delta s} = 0$}.
The only remaining free parameter in the spin-isospin channel
is  $C_1^{s}$, which is directly related to the Landau parameters
via Eqs.\ (\ref{eq:g0p}) and (\ref{eq:g1p}):
\begin{equation}
\label{eq:C1sfromg}
C_1^{s}
= \frac{1}{2 N_0} \, ( g_0' + g_1' ) ,
\end{equation}
where $g_1'$ is fixed by $C_1^T$ [also in Eq.\ (\ref{eq:g1p})].
Figure~\ref{fig:sko4p_c0p} shows results for the GT resonance energy when
$g_0'$ is systematically varied from its SkO' value by altering $C_1^s$.
We have chosen only nuclei that can be expected to exhibit a collective
response to GT excitations. Non-collective contributions may show up,
however, when the coupling constants are changed. In $^{124}$Sn,
for example, a state below the resonance collects a lot of strength for
small values of $g_0'$. Only by increasing $g_0'$ does one push that
strength into the resonance. Similarly, in $^{112}$Sn a
state about 5\,MeV above the resonance increasingly collects
strength as $g_0'$ grows.

As the underlying single-particle spectra are the same for all the cases
in Fig.~\ref{fig:sko4p_c0p}, the differences are due entirely to the
value of $g_0'$.  With increasing $g_0'$, the resonance energy increases
and more strength is pushed into the resonance.  The increase of $E_{\rm
res}$ is nearly linear, but the lines for different nuclei have
different slopes.  It is gratifying that the curves for \mbox{$E_{\rm
calc} - E_{\rm expt}$} all have a zero around the same point,
\mbox{$g_0' \approx 1.2$}.  This value is much smaller than the
empirical value \mbox{$g_0' \approx 1.8$} derived earlier
\cite{Spe77a,Ost91a,Ost92a} for at least two reasons:  (i) the
influence of the  single-particle spectrum, and (ii) the
inclusion in the residual interaction of a $p$-wave force characterized
by $g_1'$. The latter means that \mbox{$g_0'=0$} does not correspond to
a vanishing interaction in the spin-isospin channel.
%
%
\begin{figure}[t!]
\centerline{
\epsfig{file=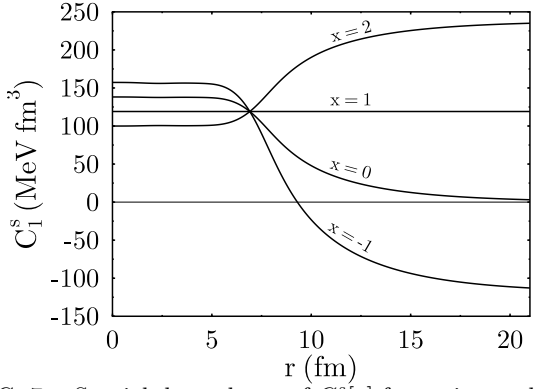}
}
\caption{\label{fig:GT:x}
Spatial dependence of $C_1^{s} [\rho]$ for various values
of $x$$\equiv$$C_1^{s}[0]$/$C_1^{s}[\rho_{\rm nm}]$,
cf.\ Eq.\ (\protect\ref{eq:ddodd}), and $g_0'$ fixed at 1.2. The value
\mbox{$x=1$} corresponds to no density dependence. For larger values
of $x$, the residual interaction becomes more repulsive outside
the nucleus than inside. When \mbox{$x=0$}, $C_1^{s}[\rho]$ vanishes
at large distances, and for negative values of $x$, the residual
interaction becomes  attractive outside the nucleus. The density profile
$\rho(r)$ used in this plot corresponds
to $^{208}$Pb.
}
\end{figure}
%
%
%
%
\subsubsection{Study of $C_1^{s} [0]$.} 
Thus far we have chosen not to let $C_1^{s}$ depend
on the density. Little is known about the empirical density dependence
of the time-odd energy functional, and time-odd Landau parameters
calculated from a ``realistic'' one-boson exchange potential in DBHF
show only a very weak density dependence \cite{Bro90a}. Because the
kinetic spin term \mbox{$C_1^T \vec{s}_{1t_3} \cdot \vec{T}_{1t_3}$},
when evaluated in INM, also contributes to the density dependence
of the Landau parameters, the density-dependence of that term must
either be small or nearly canceled by other time-odd terms.
In any event, in the following, we investigate what happens when $C_1^{s}$
depends on the (isoscalar) density in the  ``standard" way  (\ref{eq:ddodd}).
All nuclei we look at have finite neutron excess, which means that
the central density should be slightly smaller than $\rho_{\rm nm}$.
%
%
\begin{figure}[t!]
\epsfig{file=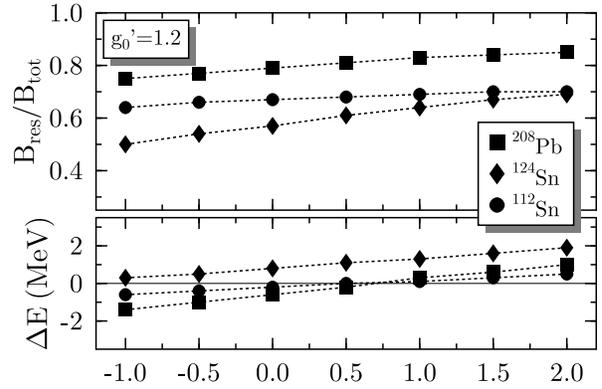}
\caption{\label{fig:sko4p_c3p}
Variation of the GT resonance energy and the strength
in the resonance when the ratio
$x$$\equiv$$C_1^{s}[0]$/$C_1^{s}[\rho_{\rm nm}]$ of parameters
defining the density dependence of $C_1^{s} [\rho]$ in
Eq.\ (\protect\ref{eq:ddodd}) is varied. Symbols and scales
are as in Fig.~\protect\ref{fig:sko4p_c0p}.
}
\end{figure}
%
%

If $g_0'$ and $g_1'$ are fixed in saturated INM, there is one free
parameter, $C_1^{s}[0]$, with which one can vary the density
dependence (\ref{eq:ddodd}). ($C_1^{s} [\rho_{\rm nm}]$ is
fixed by the value $g_0'[\rho_{\rm nm}]$=1.2, and we
set the exponent $\xi$=0.25, as it is in the time-even energy
functional SkO'.)  We continue here to assume that gauge invariance
holds, and that \mbox{$C_1^{\Delta s} = 0$}.

We vary the parameter $C_1^{s}[0]$ between $-C_1^{s}[\rho_{\rm nm}]$ and
$2C_1^{s}[\rho_{\rm nm}]$.  Figure~\ref{fig:GT:x} shows the spatial
dependence of $C_1^{s} [\rho]$ for several values of the ratio
$x$$\equiv$$C_1^{s}[0]$/$C_1^{s}[\rho_{\rm nm}]$. By changing
$C_1^{s}[0]$, one can change both the GT resonance energy and the amount
of the low-lying strength, even with $g_0' [\rho_{\rm nm}]$ kept
constant.  As Fig.\ \ref{fig:sko4p_c3p} shows, an increase of
$C_1^{s}[0]$  for a given $g_0'$ has almost the same effect as an
increase of $g_0'$ for a given $C_1^{s}[0]$. Thus, the INM Landau
parameters do not tell the whole story in a finite nucleus.
Figures~\ref{fig:GT:x} and \ref{fig:sko4p_c3p} show that the spin-spin
coupling has the largest effect on the GT resonance when it is located
at or even slightly outside the nuclear radius.
%
%
\subsubsection{Study of $C_1^{\Delta s}$.} 
%
%
\begin{figure}[b!]
\epsfig{file=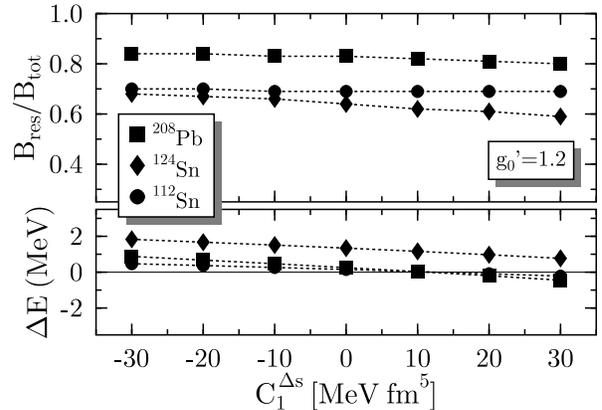}
\caption{\label{fig:sko4p_c2p}
Variation of the GT resonance energy and the strength
in the resonance when $C_1^{\Delta s}$ is varied.
Symbols and scales are as in Fig.~\protect\ref{fig:sko4p_c0p}.
}
\end{figure}
%
%
The term \mbox{$C_1^{\Delta s} \vec{s}_{1t_3} \cdot \Delta
\vec{s}_{1t_3}$} is sensitive to spatial variations of the isovector
spin density.  Unlike its (isoscalar) time-even counterpart
\mbox{$C_0^{\Delta \rho} \rho_0 \Delta \rho_0$}, it should not be called
a ``surface term'' because the spatial distribution of $\vec{s}_1$ is
determined by a few single-particle states that do not necessarily vary
the most at the nuclear surface.  In discussing the effects of this term,
we continue to fix $C_1^T$ at its Skyrme-force value via gauge
invariance and choose $C_1^s$ to be density-independent and fixed from
Eq.\ (\ref{eq:C1sfromg}) with \mbox{$g_0' [\rho_{\rm nm}] = 1.2$}.  We
then vary $C_1^{\Delta s}$ over the range of $\pm 30 \, \MeV \, \fm^5$,
covering the values obtained from the original Skyrme forces.  As seen
in Fig.\ \ref{fig:sko4p_c2p}, an increase of \mbox{$C_1^{\Delta s}$ by
$30 \, \MeV \, \fm^5$} has nearly the same effect on the GT resonance
energies as a decrease of $g_0'$ by 0.2, again demonstrating that the
value of $g_0'$ does not completely characterize the residual
interaction in finite nuclei.  A new feature of $C_1^{\Delta s}$,
apparent from the curves for $^{112}$Sn and $^{208}$Pb in Fig.\
\ref{fig:sko4p_c2p}, is the ability to move the resonance around in
energy without changing its strength.

%
%
\subsubsection{Study of $C_1^{T}$.} 
%
%
\begin{figure}[b!]
\epsfig{file=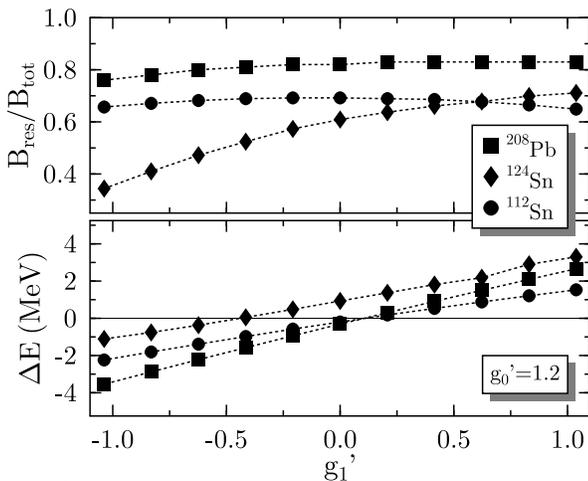}
\caption{\label{fig:sko4p_c1p}
Variation of the GT resonance energy and the strength in the
resonance when $C_1^T$ (and thus $g_1'$) is varied. $C_1^s$
is readjusted for each value of $C_1^T$ so that \mbox{$g_0' = 1.2$}.
Symbols and scales are as in Fig.~\protect\ref{fig:sko4p_c0p}.
}
\end{figure}
%
%
Finally, we investigate the influence on the GT strength distribution of
the term \mbox{$C_1^T \vec{s}_{1t_3} \cdot \vec{T}_{1t_3}$}, which
determines $g_1'$ [see Eq.\ (\ref{eq:g1p})].  As this term is linked by
gauge invariance (\ref{eq:gauge}) to the time-even $\tensor{J}^2_1$
term, a fully self-consistent variation of $C_1^T$ would require
refitting the whole time-even sector of the Skyrme functional.  (Note
that our approach removes the constraints (\ref{tx2BC}) that link
$C_t^T$ to the time-even coupling constants $C_t^\tau$ and $C_t^{\Delta
\rho}$.  The constraint was retained, however, when Sk0' was
constructed.)  We leave that task for the future, using a
gauge-invariance breaking-energy functional here with \mbox{$C_1^T \neq
C_1^J$} to obtain constraints on $C_1^T$ for future fits.
Figure~\ref{fig:sko4p_c1p} shows the change in the GT resonance when
$g_1'$ is varied in the range \mbox{$-1 \leq g_1' \leq 1$}. Increasing
$g_1'$ increases the energy of the GT resonance for a given $g_0'$.
Changing $g_1'$ by 0.2 has nearly the same effect on the GT resonance
energy as changing $g_0'$ by 0.2. (This means that \mbox{$g_0'=1.2$},
\mbox{$g_1'=0.2$}, as used here, is consistent with the lower end of
the values \mbox{$1.4 \leq g_0' \leq 1.6$}, \mbox{$g_1'=0.0$} given in
\cite{Ber81a,Gaa81aE,Suz82a,Ber81b}.) As the curves for $^{208}$Pb and
$^{112}$Sn demonstrate, however, the amount of strength in the resonance
does not necessarily change when $g_1'$ is varied.
%
%
\subsection{Regression analysis of the GT resonances}
\label{Subsect:GT:reg}
In the previous subsection we explored the dependence of the
GT resonance energies and strengths on particular time-odd
coupling constants of the Skyrme functional while keeping
the other coupling constants fixed. These results show that the GT
properties depend on all the coupling constants simultaneously, and
the effect of varying one coupling constant may be either enhanced or
cancelled by a variation of another one. In such a situation,
linear-regression is needed to quantify the influence of the coupling
constants.

We analyze the situation by supposing that the GT energies and strengths
are linear functions of four coupling constants, i.e.,
\begin{mathletters}
\label{eq:reg}
\begin{eqnarray}
\label{eq:reg-e}
E^{\rm GT}_{\rm reg}
\!& = &   e_0
       + e_1 C_1^{s}[0]
       + e_2 C_1^{s}[\rho_{\rm nm}]
       + e_3 C_1^{\Delta s}
       + e_4  C_1^T ,\!\!\!
       \\
\label{eq:reg-b}
B^{\rm GT}_{\rm reg}
\!& = &   b_0
       + b_1 C_1^{s}[0]
       + b_2 C_1^{s}[\rho_{\rm nm}]
       + b_3 C_1^{\Delta s}
       + b_4  C_1^T .\!\!\!
\end{eqnarray}
\end{mathletters}
In our linear regression method, the coefficients $e_i$ and $b_i$
are determined by a least-square fit of expressions
(\ref{eq:reg-e}) and (\ref{eq:reg-b}) to the given sample of
$N$ calculated QRPA results, $E^{\rm GT}_{\rm calc}(n)$ and
$B^{\rm GT}_{\rm calc}(n)$, \mbox{$n=1,\ldots,N$}.
The calculated $B^{\rm GT}_{\rm calc}(n)$ values have been
quenched by the usual factor of 1.26$^2$.

The sample of QRPA calculations covers
the physically interesting range of values for the coupling constants.
We present here results from a sample defined by the hypercube
\begin{mathletters}
\begin{eqnarray}
\label{eq:cube-1}
g_0'                              = & 0.6(0.2)1.8, & \quad\mbox{7 values}, \\
C_1^{s}[0]/C_1^{s}[\rho_{\rm nm}] = & -1(1)2,      & \quad\mbox{4 values}, \\
C_1^{\Delta s}                    = & -40(20)40,   & \quad\mbox{5 values}, \\
C_1^T                             = & -40(10)0,    & \quad\mbox{5 values},
\end{eqnarray}
\end{mathletters}
i.e., for the sample of \mbox{$N=700$}. We use $g_0'$ instead of
$C_1^{s}[0]$ for the regression analysis to avoid combinations of
the coupling constants that leave $g_0'$ too far from 1.2, the
value advocated in Sect.~\ref{Subsect:GT:gen}.

\begin{figure}[t!]
\epsfig{file=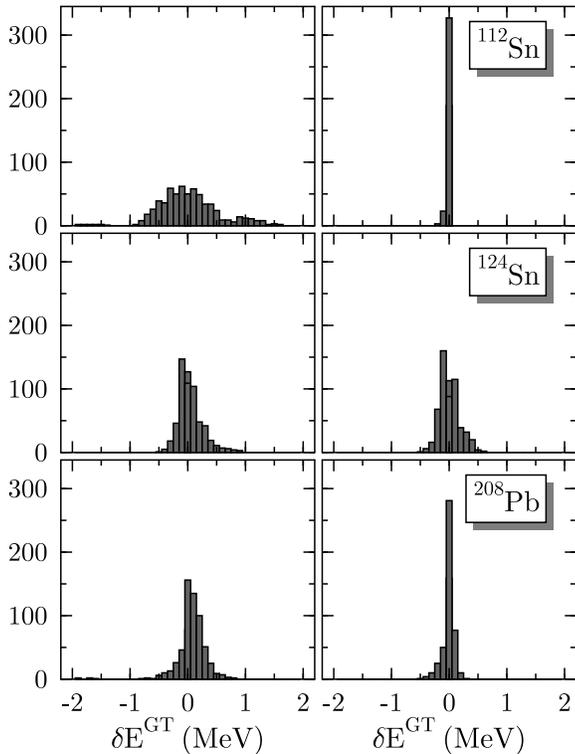}
\caption{\label{fig:his-all-e}
Distribution of differences between calculated GT resonance energies and
those from the regression analysis, with all points
from the sample (left panels) and with a reduced sample (right panels).
}
\end{figure}

\begin{figure}[t!]
\epsfig{file=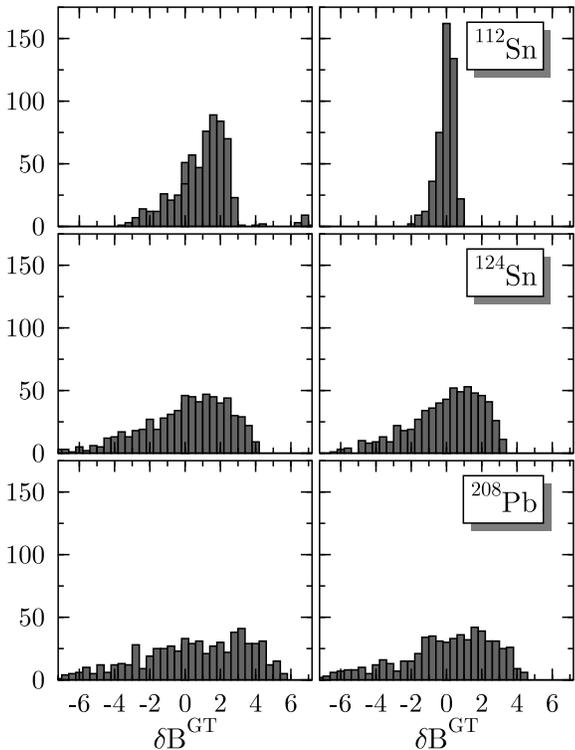}
\caption{\label{fig:his-all-s}
Same as in Fig.~\protect\ref{fig:his-all-e} except for the
strength in the GT resonance.
}
\end{figure}

The left panels in Figs.\ \ref{fig:his-all-e} and \ref{fig:his-all-s}
contain histograms of deviations
\begin{mathletters}
\begin{eqnarray}
\label{eq:dev-e}
\delta E^{\rm GT}(n)
& = &  E^{\rm GT}_{\rm calc}(n)- E^{\rm GT}_{\rm reg}, \\
\label{eq:dev-b}
\delta B^{\rm GT}(n)
& = &  B^{\rm GT}_{\rm calc}(n)- B^{\rm GT}_{\rm reg}
\end{eqnarray}
\end{mathletters}
between the calculated and fitted energies and strengths
in $^{112}$Sn, $^{124}$Sn, and $^{208}$Pb. The widths of these
distributions illustrate the degree to which the linear regression
expressions (\ref{eq:reg}) are able to describe the results of the QRPA
calculations. One can see that the fit GT resonance energies are
generally within about $\pm 1$\,MeV of the calculated ones,
and the fit strengths within about $\pm$6.
%
%
%
%
%
%
%
%

As can be seen from Figs.~\ref{fig:sko4p_c0p}, \ref{fig:sko4p_c3p},
\ref{fig:sko4p_c2p}, and \ref{fig:sko4p_c1p}, the dependence of both the
resonance energy and the strength in the resonance on the coupling
constants is not linear for the entire region of coupling constants.
Although our sample is restricted to the area around the reasonable
values, at times we leave the region where the regression can safely be
performed.  Furthermore, for certain combinations of the coupling
constants (especially at a weak coupling), there are competing states
that carry strength similar to that of the GT ``resonance.'' (These
states often merge into the resonance at a larger coupling.)  Finally,
the resonance can be fragmented into many (sometimes up to 15) states.
Therefore, we remove certain areas of parameter space where the
determination of either the energy or the strength of the GT resonance
is ambiguous.  Such areas are almost always singled out by particularly
large deviations from the fitted values.
%
%
\begin{table}[t!]
\begin{center}
\begin{tabular}{lddd}
                              & $^{112}$Sn & $^{124}$Sn &  $^{208}$Pb  \\
\hline
$e_0$                        &   5.58100  &  10.16000  &   8.19600    \\
$e_1$                        &   0.00305  &   0.00488  &   0.00674    \\
$e_2$                        &   0.02696  &   0.03981  &   0.06099    \\
$e_3$                        &   0.01690  &   0.03474  &   0.05897    \\
$e_4$                        &  -0.01189  &  -0.01767  &  -0.02198    \\
\hline
$\sigma_E$                   &   0.0479   &   0.199    &   0.114      \\
$\sigma_E$(full)             &   0.585    &   0.865    &   0.520      \\
\end{tabular}
\end{center}
\caption{\label{tab:reg-e}
Coefficients $e_i$ obtained by the regression analysis, Eq.\
(\ref{eq:reg-e}),
of the QRPA GT resonance energies in the reduced sample (see text). The
standard deviations $\sigma_E$ from this sample are compared
to the $\sigma_E$(full) from the full sample of $N$=700
points.
}
\end{table}
%
%
%
%
\begin{table}[t!]
\begin{center}
\begin{tabular}{lddd}
                              & $^{112}$Sn & $^{124}$Sn &  $^{208}$Pb  \\
\hline
$b_0$                        &  21.86000  &  11.75000  &  80.62000    \\
$b_1$                        &   0.00963  &   0.03107  &   0.03079    \\
$b_2$                        &   0.06143  &   0.25120  &   0.18620    \\
$b_3$                        &   0.09146  &   0.30510  &   0.33850    \\
$b_4$                        &  -0.01730  &  -0.10280  &  -0.09688    \\
\hline
$\sigma_B$                   &   0.608    &   2.10     &   2.74       \\
$\sigma_B$(full)             &   5.20     &   3.44     &   3.82       \\
\end{tabular}
\end{center}
\caption{\label{tab:reg-b}
Same as in Table \protect\ref{tab:reg-e} except for the strengths
of the GT resonances.
}
\end{table}
%
%

After reducing the sample in this way, we obtain the histograms in the
right panels of Figs.~\ref{fig:his-all-e}
and \ref{fig:his-all-s}.  These illustrate the quality of the
regression fits obtained for samples of \mbox{$N=542$}, 664,
and 618 in $^{112}$Sn, $^{124}$Sn, and $^{208}$Pb, respectively.
Tables \ref{tab:reg-e} and \ref{tab:reg-b} list
the corresponding values of the regression coefficients, as
well as the standard deviations for the GT energies
and strengths within each of the samples.

Figures\ \ref{fig:his-all-e} and \ref{fig:his-all-s} and the standard
deviations obtained in the reduced and full samples (Tables
\ref{tab:reg-e} and \ref{tab:reg-b}) show that the description
obtained by removing a small number of points beyond the region of
linearity is quite good.  The GT resonance energies are now reproduced
within about $\pm$200\,keV or less.  The description of the resonant GT
strengths is also improved, especially in $^{112}$Sn, although here the
linear regression cannot work too well because the strengths saturate at
strong coupling.  Nevertheless, the coefficients listed in Tables
\ref{tab:reg-e} and \ref{tab:reg-b} allow a fairly reliable estimate of
the QRPA values for any combination of the coupling constants.  The
values of the coefficients in Tables \ref{tab:reg-e} and \ref{tab:reg-b}
show that $C_1^{s}[\rho_{\rm nm}]$ and $C^{\Delta s}_1$ strongly
influence properties of the GT resonance, and that both the energies and
the resonant strengths increase when these coupling constants increase.
$C^T_1$ has a weaker effect in the opposite direction, while $C^s_1[0]$
is less important still.

Without presenting detailed results, we report here on two other
attempts at regression analysis.  We tried to analyze the
results for all the three nuclei, $^{112}$Sn, $^{124}$Sn, and
$^{208}$Pb, simultaneously by adding terms $e_5(N-Z)$ and $b_5(N-Z)$ to
the regression formulas (\ref{eq:reg}).  Linear scaling might be
obtained by analyzing the QRPA results for very many nuclei, where the
effects due to shell structure could average out.  In our small sample,
shell structure is obviously important. We also tried the regression
analysis with \mbox{$C^T_1 = -9.172$}\,MeV\,fm$^5$ fixed at its SkO'
value (see Table \ref{tab:toddcplg}), and without the terms $e_4$
and $b_4$ in the
regression formulae (\ref{eq:reg}), so that the functional's gauge
invariance was preserved.  The results were not significantly different
from those when $C^T_1$ was allowed to vary freely. Consequently,
our analysis does not allow us any constraints on $C^T_1$ that might
be used in future fits of the time-even part of the energy functional.

For the SkO' coupling constants (see Table \ref{tab:toddcplg}), we obtain
\mbox{$E^{\rm GT}_{\rm reg}=8.3$\,MeV}, 14.2\,MeV, and 14.2\,MeV in
$^{112}$Sn, $^{124}$Sn, and $^{208}$Pb, and \mbox{$B^{\rm GT}_{\rm
reg}=100$} in $^{208}$Pb.  These values are close to the corresponding
experimental data:  8.9\,MeV, 13.7\,MeV, 15.5\,MeV, from
Ref.\cite{Gaa81aE}, and \mbox{$\sim 3(N-Z)/1.26^2$}. It is not
possible, however, to find values of the four coupling constants that
reproduce these four experimental data points exactly. The reason is
that the matrix of corresponding regression coefficients is almost
singular, resulting in absurdly large values of the coupling constants.
Clearly a determination of the coupling constants from  experiment
would require more data.  Although charge-exchange measurements
have been made on many nuclei, we require spherical even-even nuclei
that are not soft against vibrations.  To fit the relevant coupling
constants to data, we would need, at the minimum, the ability to treat
deformed nuclei.  Meanwhile, we can make a simple choice of time-odd
coupling constants  from the analysis in Fig.~\ref{fig:sko4p_c0p}.  The values
\begin{eqnarray}
C_1^s [0]      & = C_1^s[\rho_{\rm nm}] = & 120 \, {\rm MeV} \, {\rm fm}^3
                  \nn \\
C_1^{\Delta s} & \phantom{C_1^s[\rho_{\rm nm}] =} = & 0
                  \nn \\
C_1^T          & \phantom{C_1^s[\rho_{\rm nm}] =} = & -9.172 \, {\rm 
MeV} \, {\rm fm}^5
\end{eqnarray}
(see Table IV) give $g_0'=1.2$ and $g_1'$ = 0.19, which are in accord
with the data we discuss.  But these values by no means constitute a fit
and are not unique.
%
%
\begin{figure*}[t!]
\centerline{\epsfig{file=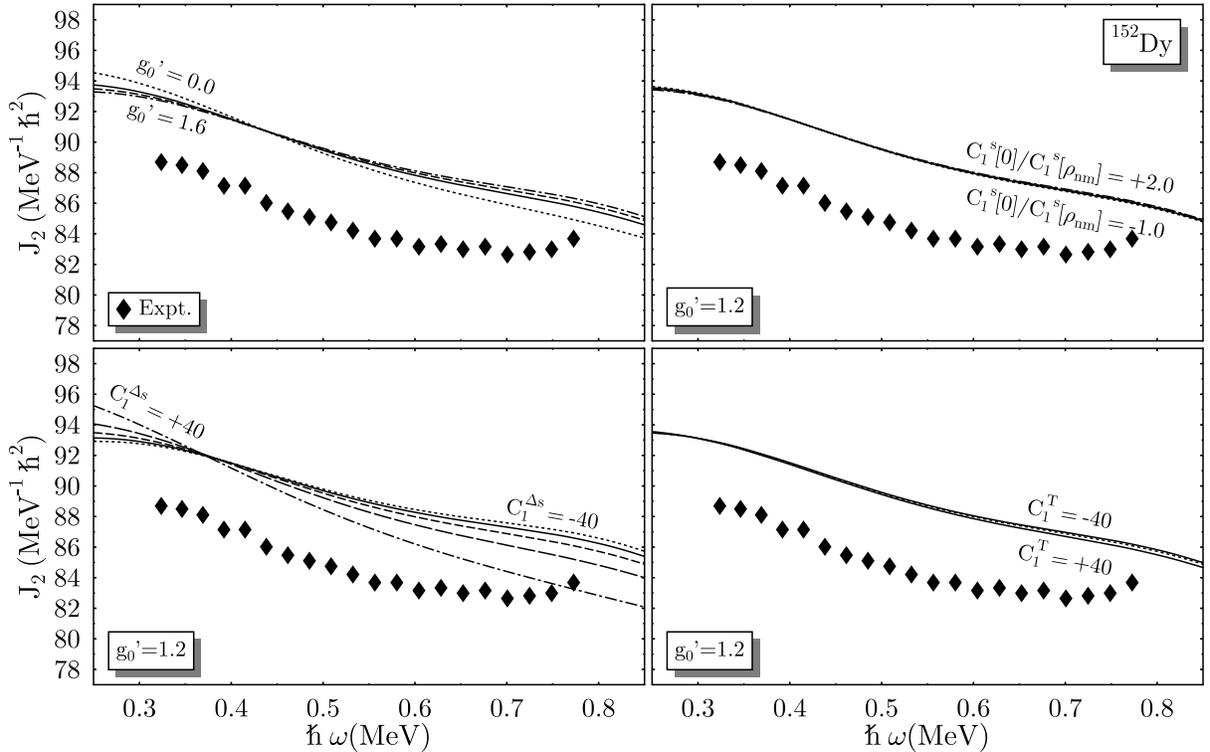}}
\caption{\label{Fig:rot:skop}
Dynamical moment of inertia $\Jt$ in the superdeformed band of
$^{152}$Dy calculated with the SkO' energy density functional and
modified time-odd coupling constants.
In the upper left panel (corresponding to Fig.~\protect\ref{fig:sko4p_c0p})
all coupling constants are chosen to be density independent, $C_1^{\tau}$
is kept at the Skyrme-force value and \mbox{$C_1^{\Delta s}=0$}.
In the upper right panel (corresponding to Fig.~\protect\ref{fig:sko4p_c3p})
the density dependence of $C_1^{s}$ is varied keeping \mbox{$g_0' = 1.2$}.
In the lower left panel (corresponding to Fig.~\protect\ref{fig:sko4p_c2p})
$C_1^{\Delta s}$ is varied, while in the lower right panel
(corresponding to Fig.~\protect\ref{fig:sko4p_c1p}) $C_t^{T}$ is varied.
See text for the choice of isoscalar time-odd couplings.
}
\end{figure*}
%
%

%
%
\section{A consistency check: Superdeformed rotational bands}
\label{Sect:SD}
Another phenomenon in which the time-odd part of the Skyrme energy
density functional plays a role is the high-spin rotation of very
elongated nuclei. In this section we demonstrate that reasonable
values for the spin-isospin coupling constants found when analyzing
the GT strength are consistent with the description of superdeformed
rotational bands.

When a nucleus rotates rapidly, there appear strong current and spin
one-body densities along with the usual particle densities that
characterize stationary (time-even) states.  The time-odd densities are
at the origin of strong time-odd mean fields.   There are already many
self-consistent studies of high-spin states available; see, e.g., reviews
in Refs.~\cite{Rin96,Abe90a,Jan91,Bak95}. The role and significance of
the time-odd mean-field terms, however, has not been carefully studied.
Basic features of high-spin states can often be well described by models
that use phenomenological mean fields of the Woods-Saxon or Nilsson
type, where no time-odd terms are explicitly present in the one-body
potential. (The time-odd densities are, however, present there through
the time-odd cranking term.) For the Gogny interaction \cite{Dec80a}, or
within the standard RMF models \cite{Rin96}, they cannot be
independently modified; the Gogny interaction is defined as a two-body
force (where the time-odd terms show up as exchange terms), while all
time-odd terms appearing in standard RMF models are fixed by Lorentz
invariance. Within the Skyrme framework, the time-odd terms in
superdeformed rotational states were analyzed in an exploratory
way in Refs.\ \cite{Dob95c,Dob96d}.

Unlike the GT response, rotational bands are influenced by both
isoscalar and isovector time-odd channels of the effective
interaction. In fact, the large effects of time-odd coupling constants
found in
\cite{Dob95c} are mainly due to the isoscalar channel; the
isovector channel induces corrections that are smaller, though
non-negligible.  The SkO' Skyrme parameterization,
which we use for GT calculations,
is unstable when the original parameters from
Eq.\ (\ref{eq:cpl:SF}) are used in the isoscalar spin channel because
\mbox{$g_0 < -1$} (a fact that is related
to the unusually high value of $g_1$ in Table~\ref{tab:landau}).
This leads to unphysical ferromagnetic solutions where all
spins align when the nucleus is cranked.
Of course, the value of $g_0$ does not influence the GT calculations
for even-even nuclei presented in our study which focuses on
  the isovector  time-odd coupling constants. Consequently, in the following
we employ a simple spin energy functional using the
Skyrme force value for $C_0^T$, setting \mbox{$C_0^{\Delta s} = 0$}
and neglecting density dependence. We adopt the value \mbox{$g_0 = 0.4$}
given in \cite{Ost92a} (note that a different definition of the
normalization factor is used there) to fix $C_0^s$.

We perform the calculations in exactly the same way as
in Ref.\ \cite{Dob95c} by using the code HFODD (v1.75r) described
in Ref.\cite{Dob00d}. We examine $^{152}$Dy, which is a doubly magic
superdeformed system. Pairing has a minor influence and we neglect
it. We focus on the dynamic moment of inertia ${\cal J}_2$:
\begin{equation}
\Jt (I)
   =  \left[ \frac{d^2E}{dI^2} \right]^{-1}
   \simeq  \frac{4\hbar^2 }{\Delta E_\gamma}
\end{equation}
(from experimental data) or
\begin{equation}
\Jt (\omega_1)
   =   \frac{dI}{d \omega}
   \simeq   \frac{I(\omega_1) - I(\omega_2)}{\omega_1 - \omega_2}
\end{equation}
(in calculations). Figure\ \ref{Fig:rot:skop} shows results of
calculations when one of the four time-odd isovector coupling constants
is varied, while the other ones are kept at the values mentioned above.
Variations of the coupling constants $C_1^{s}[0]$,
$C_1^{s}[\rho_{\rm nm}]$, and $C_1^{T}$ have little effect on
the dynamic moments of inertia in $^{152}$Dy.  When $C_1^{\Delta s}$ is
varied, the moments change noticeably, but the general trend with
frequency is still the same.  Thus, altering the isovector time-odd
couplings does not appear to change the quality with which we describe
superdeformed rotational bands.  Of course,
a consistent description of both the high-spin data and the GT
resonance properties over a wide range of nuclei will require a much
more detailed analysis.
%
%
\section{Summary, conclusions, and outlook}
\label{Sect:SCO}
By exploiting the freedom in the Skyrme energy functional, we have taken
significant steps towards a fully self-consistent description of nuclear
ground states and the GT response.  Along the way, we debunked the
notion that the strength and location of the GT resonance in finite nuclei
is determined entirely by the Landau parameter $g_0'$. Our analysis also
shows this parameter to be smaller than previous work indicates.

There are not enough experimental data for spherical even-even nuclei
to fix the time-odd isovector
coupling constants; the ability to do calculations in deformed nuclei
should help there. We could, however, choose values that reproduce the
data we do analyze, without spoiling our description of high-spin
superdeformation. Doing a lot better may require improving our time-even
energy functionals. GT resonance energies and strengths depend
significantly on spin-orbit splitting as well as the residual
spin-isospin interaction. Until we are better able to reproduce
single-particle energies, therefore, a fit of the time-odd interaction
will be tentative.

We have not considered isoscalar time-odd interactions. The couplings
there will be harder to fix because there are fewer data on the
response, which is not as collective as in the charge-exchange channel.
In addition, the isovector time-odd terms will play a role in
calculations of isoscalar observables. Though a lot clearly remains to
be done, our work can already be put to good use.  We will, for example,
employ the new values for the isovector time-odd coupling constants in
future calculations of beta decay and in the observables that tell us
about the extent of real time-reversal violation in nuclei.
%
%
\acknowledgements
This work was supported in part by the U.S.\ Department of Energy
under Contract Nos.\ DE-FG02-96ER40963 (University of Tennessee),
DE-FG02-97ER41019 (University of North Carolina), DE-AC05-00OR22725
with UT-Battelle, LLC (Oak Ridge National Laboratory), DE-FG05-87ER40361
(Joint Institute for Heavy Ion Research), by the Polish Committee for 
Scientific Research (KBN)  under Contract No.\ 5~P03B~014~21, and by the
Wallonie/Brussels-Poland integrated actions program. We thank the
Institute for Nuclear Theory at the University of Washington for its
hospitality during the completion of this work.
%
%
\begin{appendix}
\section{Local Densities and Currents}
\label{Sect:app:dens}
The complete density matrix $\rho(\rvec \sigma t, \rvecp \sigma' t')$
in spin--isospin space as defined in (\ref{eq:densitymatrix})
can be decomposed into the sum of scalar $\rho_{t t_3} (\rvec, \rvecp)$
and vector densities $\vec{s}_{t t_3} (\rvec, \rvecp)$, where the
subscripts
denote the isospin quantum numbers:
\begin{eqnarray}
\lefteqn{\rho(\rvec \sigma \tau, \rvecp \sigma' \tau')
} \nn \\
& = & \tfrac{1}{4} \Big[
       \rho_{00} (\rvec, \rvecp) \, \delta_{\sigma \sigma'} \,
       \delta_{\tau \tau'}
     + \vec{s}_{00} (\rvec, \rvecp) \cdot \sigmavec_{\sigma \sigma'}
       \, \delta_{\tau \tau'}
       \nonumber \\
&   & + \delta_{\sigma \sigma'} \! \sum_{t_3 = -1}^{+1}
       \rho_{1 t_3} (\rvec, \rvecp) \tau^{t_3}_{\tau\tau'}
     +  \sum_{t_3 = -1}^{+1}
       \vec{s}_{1 t_3} (\rvec, \rvecp) \cdot \sigmavec_{\sigma \sigma'}
       \tau^{t_3}_{\tau\tau'}
   \Big]
\nn \\
\end{eqnarray}
The quantities $\sigmavec_{\sigma \sigma'}$ and $\tau^{t_3}_{\tau\tau'}$ are
matrix elements of the Pauli matrices in spin and isospin space.
In terms of these, the local density $\rho$, spin density $\vec{s}$,
kinetic density $\tau$, kinetic spin density $\vec{T}$, current $\vec{j}$,
and spin-orbit tensor $\tensor{J}$ are
\begin{eqnarray}
\rho_{t t_3} (\rvec)
& = & \rho_{t t_3} (\rvec, \rvec) \phantom{\Big|_{\rvec=\rvecp}} \nn \\
\vec{s}_{t t_3} (\rvec)
& = & \vec{s}_{t t_3} (\rvec, \rvec) \phantom{\Big|_{\rvec=\rvecp}} \nn
\\
\tau_{t t_3} (\rvec)
& = & \nabla \cdot \nabla' \rho_{t t_3} (\rvec,\rvecp)
       \Big|_{\rvec=\rvecp} \nn \\
\vec{T}_{t t_3} (\rvec)
& = & \nabla \cdot \nabla'
       \vec{s}_{t t_3} (\rvec,\rvecp) \Big|_{\rvec=\rvecp} \nn \\
\vec{j}_{t t_3} (\rvec)
& = & - \tfrac{\iunit}{2} ( \nabla - \nabla' )
         \rho_{t t_3} (\rvec,\rvecp) \Big|_{\rvec=\rvecp} \nn \\
J_{t t_3, ij} (\rvec)
& = & - \tfrac{\iunit}{2} ( \nabla - \nabla' )_i \;
       s_{t t_3,j} (\rvec,\rvecp) \Big|_{\rvec=\rvecp}
       \quad .
\end{eqnarray}
The densities $\rho$, $\tau$, and $\tensor{J}$ are time-even, while
$\vec{s}$,  $\vec{T}$, and $\vec{j}$ are time-odd. See \cite{Dob96d}
for a more detailed discussion.
%
%
\section{Energy density functional from the two-body Skyrme force}
\label{Sect:app:SF}
%
%
\begin{table*}[t]
\caption{\label{tab:toddcplg}
Time-odd coupling constants calculated from Eq.\
(\protect\ref{eq:cpl:SF})
for the Skyrme interactions as indicated.
}
\begin{center}
\begin{tabular}{lddddddddc}
\hline\noalign{\smallskip}
Force & $C_0^{s}[0]$
       & $C_1^{s}[0]$
       & $C_0^{s}[\rho_{\rm nm}]$
       & $C_1^{s}[\rho_{\rm nm}]$
       & $C_0^{T}$
       & $C_1^{T}$
       & $C_0^{\Delta s}$
       & $C_1^{\Delta s}$
       & $\alpha$ \\
       & $(\MeV \, \fm^3)$
       & $(\MeV \, \fm^3)$
       & $(\MeV \, \fm^3)$
       & $(\MeV \, \fm^3)$
       & $(\MeV \, \fm^5)$
       & $(\MeV \, \fm^5)$
       & $(\MeV \, \fm^5)$
       & $(\MeV \, \fm^5)$
       & \\
\noalign{\smallskip}\hline\noalign{\smallskip}
SkI1  &  695.860 & 239.200 &  120.190 & 99.573  &    0.0   &   0.0   &
192.660 & 62.766 & $1/4$ \\
SkI3  &   84.486 & 220.360 &  253.180 & 113.940 &    0.0   &   0.0   &
92.235 & 22.777 & $1/4$ \\
SkI4  &   44.038 & 231.980 &  209.030 & 104.120 &    0.0   &   0.0   &
124.590 & 37.943 & $1/4$ \\
SkO   &  373.770 & 262.960 &   41.421 &  84.253 &    0.0   &   0.0   &
70.365 & 26.590 & $1/4$ \\
SkO'  &  277.910 & 262.430 &   47.082 &  84.154 & -104.090 &  -9.172 &
42.791 & 16.553 & $1/4$ \\
SkX   &   57.812 & 180.660 &  -35.639 &  81.246 &   -7.861 & -23.669 &
-4.434 &  9.514 & $1/2$ \\
SGII  &  271.110 & 330.620 &   61.048 &  91.676 &    0.0   &   0.0   &
15.291 & 15.283 & $1/6$ \\
SkP   &  152.340 & 366.460 &  -31.328 &  78.562 &    7.713 & -41.127 &
-4.211 &  9.757 & $1/6$ \\
SkM*  &  271.110 & 330.620 &   31.674 &  91.187 &    0.0   &   0.0   &
17.109 & 17.109 & $1/6$ \\
SLy4  & -207.820 & 311.110 &  153.210 &  99.737 &    0.0   &   0.0   &
47.048 & 14.282 & $1/6$ \\
SLy5  & -171.360 & 310.430 &  151.080 &  99.133 &  -14.659 & -65.058 &
45.787 & 14.000 & $1/6$ \\
SLy6  & -201.460 & 309.940 &  157.050 & 100.280 &    0.0   &   0.0   &
48.822 & 14.655 & $1/6$ \\
SLy7  & -215.830 & 310.100 &  158.260 & 100.640 &  -30.079 & -55.951 &
49.680 & 14.843 & $1/6$ \\
\noalign{\smallskip}\hline\noalign{\smallskip}
\end{tabular}
\end{center}
\end{table*}
%
%
The standard two-body Skyrme force is given by \cite{Eng75a,Vau72a}
\begin{eqnarray}
\label{Skyrmeforce}
\lefteqn{v_{\rm Skyrme} (\vec{r}_1, \vec{r}_2)}
       \nn \\
& = & t_0 \, ( 1 + x_0 \Pop_\sigma ) \;
       \delta (\rvec_1 - \rvec_2)
       \nn \\
&   &
       + \tfrac{1}{2} \; t_1 \; ( 1 + x_1 \Pop_\sigma )
         \Big[   \hat{\vec{k}}{}^{\prime2} \;
                 \delta (\rvecp_1 - \rvecp_2)
               + \delta (\rvec_1  - \rvec_2)
                 \; \hat{\vec{k}}{}^2 \Big]
       \nn \\
&   &
       + t_2 \, ( 1 + x_2 \Pop_\sigma ) \;
         \hat{\vec{k}}{}' \cdot
         \delta (\rvec_1  - \rvec_2) \;
         \hat{\vec{k}}
       \nn \\
&   &
       + \tfrac{1}{6} \, t_3 \; ( 1 + x_3 \Pop_\sigma ) \;
         \delta (\rvec_1  - \rvec_2) \;
         \rho^\alpha \left( \tfrac{\rvec_1 + \rvec_2}{2} \right)
       \nn \\
&   &
       + \iunit W_0 \,
         ( \hat{\sigmavec}_1 + \hat{\sigmavec}_2 ) \cdot
         \hat{\vec{k}}{}' \times
         \delta (\rvec_1 - \rvec_2) \;
         \hat{\vec{k}}
\quad ,
\end{eqnarray}
where \mbox{$\Pop_\sigma = \half(1 + \hat{\sigmavec}_1 \cdot
\hat{\sigmavec}_2$)} is the spin-exchange operator,
\mbox{$\hat{\vec{k}} = - \tfrac{\iunit}{2} (\nabla_1 - \nabla_2)$}
acts to the right, and
\mbox{$\hat{\vec{k}}{}' = \tfrac{\iunit}{2} (\nabla_1' - \nabla_2')$}
acts to the left. Calculating the Hartree-Fock expectation value from
this force yields the energy functional given in Eq.\
(\ref{eq:SkyrmeFu:gauge}) with the coupling constants:
\begin{eqnarray}
\label{eq:cpl:SF}
\label{tx2BC}
C_0^\rho  & = & \tfrac{3}{8}  t_0 + \tfrac{3}{48} t_3 \, \rho_0^\alpha
                 \nn \\
C_1^\rho  & = & - \tfrac{1}{4}  t_0 \Big( \half + x_0 \Big)
                 - \tfrac{1}{24} t_3 \Big( \tfrac{1}{2} + x_3 \Big)
                   \, \rho_0^\alpha
                 \nn \\
C_0^s     & = & - \tfrac{1}{4} t_0 \Big( \half - x_0 \Big)
                 - \tfrac{1}{24} t_3 \Big( \tfrac{1}{2} - x_3 \Big)
                   \, \rho_0^\alpha
                 \nn \\
C_1^s     & = & - \tfrac{1}{8} t_0
                 - \tfrac{1}{48} t_3  \, \rho_0^\alpha
                 \nn \\
C_0^\tau  & = &   \tfrac{3}{16} \, t_1
                 + \tfrac{1}{4} t_2 \; \Big( \tfrac{5}{4} + x_2 \Big)
                 \nn \\
C_1^\tau  & = & - \tfrac{1}{8} t_1 \Big(\tfrac{1}{2} + x_1 \Big)
                 + \tfrac{1}{8} t_2 \Big(\tfrac{1}{2} + x_2 \Big)
                 \nn \\
C_0^T     & = & \switchJ \,
                 \Big[
                 - \tfrac{1}{8} t_1 \Big( \tfrac{1}{2} - x_1 \Big) \,
                 + \tfrac{1}{8} t_2 \Big( \tfrac{1}{2} + x_2 \Big)
                 \Big]
                 \nn \\
C_1^T
& = & \switchJ \,
       \Big[ - \tfrac{1}{16} t_1
             + \tfrac{1}{16} t_2
       \Big]
       \nn \\
C_0^{\Delta \rho}
& = & - \tfrac{9}{64} t_1
       + \tfrac{1}{16}  t_2 \Big( \tfrac{5}{4} + x_2 \Big)
       \nn \\
C_1^{\Delta \rho}
& = &   \tfrac{3}{32} t_1 \Big( \tfrac{1}{2} + x_1 \Big)
       + \tfrac{1}{32} t_2 \Big( \tfrac{1}{2} + x_2 \Big)
       \nn \\
C_0^{\Delta s}
& = &   \tfrac{3}{32} t_1 \Big( \tfrac{1}{2} - x_1 \Big)
       + \tfrac{1}{32} t_2 \Big( \tfrac{1}{2} + x_2 \Big)
       \nn \\
C_1^{\Delta s}
& = & \tfrac{3}{64} t_1 + \tfrac{1}{64} t_2
       \nn \\
C_0^{\nabla J}
& = & - \tfrac{3}{4} W_0
       \nn \\
C_1^{\nabla J}
& = & - \tfrac{1}{4} W_0
       \nn \\
C_0^{\nabla s}
& = & 0
       \nn \\
C_1^{\nabla s}
& = & 0~,
\end{eqnarray}
nine of which are independent. Although in this approach
\mbox{$\switchJ = 1$}, many  parameterizations of the Skyrme
interaction set \mbox{$\switchJ = 0$}. That violates the interpretation
of the Skyrme functional as an expectation value of a real two-body
interaction and removes the rationale for calculating the time-odd
coupling constants from (\ref{tx2BC}). For Skyrme interactions with a
generalized spin-orbit interaction \cite{Rei95a}, e.g.\ for SkI3,
SkI4, SkO, or SkO', the spin-orbit coupling constants are given by
\begin{equation}
C_0^{\nabla J}
= - b_4 - \half b_4'
\qquad , \qquad
C_1^{\nabla J}
= - \half b_4'
\quad .
\end{equation}
The resulting terms in the energy functional again cannot be represented
as the HF expectation value of a two-body spin-orbit potential (see,
e.g.,\ \cite{Sha95a}), again violating the assumptions behind the
calculation of the time-odd coupling constants in (\ref{tx2BC}).

As Eqs.~(\ref{tx2BC}) represent the standard approach to the time-odd
coupling constants, it is worthwhile to take a look at the actual
values. Table \ref{tab:toddcplg} compares them for several Skyrme
forces. None of these parameterizations was obtained from observables 
sensitive to the time-odd terms in the energy functional. Differences 
among the forces merely reflect various strategies for adjusting the 
time-even coupling constants. Values of the density-dependent isoscalar 
coupling constants $C_0^{s}$, either at
\mbox{$\rho_0 = 0$} or at \mbox{$\rho_0 = \rho_{\rm nm}$}, are scattered
in a wide range. This is probably one of the main sources of
differences in the predictions of the forces for time-odd corrections
to rotational bands. For the SLy$x$ forces, $C_0^{s} [0]$
is negative, which is unusual; most often
this part of the isoscalar spin-spin interaction is repulsive at all
densities. The difference will probably cause visible differences
in rotational properties whenever the spin density is large at the
surface. All the forces agree on the isovector coupling constant
$C_1^{s}$, especially at the saturation density, i.e.,
\mbox{$C_1^s[\rho_{\rm nm}] \approx 100 \;$MeV fm$^3$}. This
simply follows from the fact that, assuming Eq.\ (\ref{eq:cpl:SF}),
$C_1^{s}$ is proportional to the time-even $C_0^{\rho}$ that is
fixed from binding energies and radii.
%
%
\section{Infinite Nuclear Matter}
\label{Sect:app:INM}
\subsection{Introduction}
Homogeneous infinite nuclear matter (INM) is  widely used to study
and characterize nuclear interactions. Some INM properties,
such as the saturation density, energy per particle, and
asymmetry coefficient, are coherent, and others, such as the
incompressibility $K_\infty$ and the sum-rule enhancement factor,
are related to excitations and can be used as
pseudo-observables to compare with predictions of nuclear forces.
INM properties are also often used to adjust the parameters
of effective interactions for self-consistent calculations.
These properties at large asymmetry are key ingredients for
the description of neutron stars. (See, e.g.,\ Refs.\ \cite{Cha97a,Hae89a}
for a discussion on the mean-field level.)

Most papers deal with spin-saturated INM, in which the time-odd
channels of the interaction discussed here do not contribute.
Nothing is known about spin-polarized INM, which
actually may play some role in neutron stars.
A stability criterion for this exotic system, derived in
Ref.\ \cite{Kut94a}, was even used to adjust the parameters
of the SLy$x$ forces in Ref.\cite{Cha97a,Cha98a}.
We do not consider tensor forces in this work. Their contribution
to the properties of polarized INM were explored, e.g., in
Ref.~\cite{Hae82a}.
%
%
\subsection{Degrees of freedom}
The four basic degrees of freedom of homogeneous INM are the
isoscalar scalar density
$\rho_0$, the isovector scalar density $\rho_1$, the isoscalar
vector density $s_0$, and the isovector vector density $s_1$.
They can be expressed through the usual neutron and proton,
spin-up, and spin-down densities in the following way.
\begin{eqnarray}
\rho_0
& = & \rhonu + \rhond + \rhopu + \rhopd ,
       \nn \\
\rho_1
& = & \rhonu + \rhond - \rhopu - \rhopd ,
       \nn \\
s_0
& = & \rhonu - \rhond + \rhopu - \rhopd ,
       \nn \\
s_1
& = & \rhonu - \rhond - \rhopu + \rhopd .
\end{eqnarray}
Similarly, densities of protons and neutrons
with spin up and down can be expressed as:
\begin{eqnarray}
\rhonu
& = & \tfrac{1}{4} ( \rho_0 + \rho_1 + s_0 + s_1 )
   =   \tfrac{1}{4} ( 1 + I_\tau + I_\sigma + I_{\tau \sigma} ) \; \rho_0
,
       \nn \\
\rhond
& = & \tfrac{1}{4} ( \rho_0 + \rho_1 - s_0 - s_1 )
   =   \tfrac{1}{4} ( 1 + I_\tau - I_\sigma - I_{\tau \sigma} ) \; \rho_0
,
       \nn \\
\rhopu
& = & \tfrac{1}{4} ( \rho_0 - \rho_1 + s_0 - s_1 )
   =   \tfrac{1}{4} ( 1 - I_\tau + I_\sigma - I_{\tau \sigma} ) \; \rho_0
,
       \nn \\
\rhopd
& = & \tfrac{1}{4} ( \rho_0 - \rho_1 - s_0 + s_1 )
   =   \tfrac{1}{4} ( 1 - I_\tau - I_\sigma + I_{\tau \sigma} ) \; \rho_0
,
       \nn \\
\end{eqnarray}
where
$I_\tau          = \rho_1 / \rho_0$ is the relative isospin excess,
$I_\sigma        = s_0    / \rho_0$ is the relative spin excess, and
$I_{\sigma \tau} = s_1    / \rho_0$ is the relative spin-isospin excess,
with $-1 \leq I_i \leq +1$.

In symmetric unpolarized INM $I_i = 0$, while in asymmetric INM
$\rho_1 \neq 0$. Polarized INM has $s_0 \neq 0$, and
spin-isospin polarized nuclear matter has $s_1 \neq 0$.
%
%
\subsection{Fermi surfaces and kinetic densities}
For INM arbitrary asymmetry, the Fermi energy of each
particle species is different. Finite spin densities $s_i$ break the
isotropy of INM, creating the possibility that the Fermi surface will
deform\cite{Dab72a}. We are mainly interested in INM with small
polarization, so we use the approximation that all Fermi surfaces
are spherical.

In the mean-field approximation, $\tilde\rho_{q \sigma} (\kvec)$, the
density of particles in momentum space with the isospin projection $q$
and spin projection $\sigma$, is
\begin{equation}
\tilde\rho_{q \sigma} (\kvec)
= \left\{
\begin{array}{l}
           1 \quad \mbox{for $k \leq k_{{\rm F}, q \sigma}$}, \\
           0 \quad \mbox{for $k >    k_{{\rm F}, q \sigma}$}.
           \end{array}
   \right.
\end{equation}
In asymmetric polarized INM, the relation between the Fermi momenta
and the isoscalar scalar density reads
\begin{equation}
\rho_0
= \frac{2}{3 \pi^2} \; k_{\rm F}^3
= \frac{1}{6 \pi^2}
   \sum_{q = p,n}
   \sum_{\sigma = \uparrow, \downarrow}
   k_{{\rm F}, q \sigma}^3
   \quad ,
\end{equation}
with $k_{{\rm F}, q \sigma} = (6 \pi^2)^{1/3} \rho_{q \sigma}^{1/3}$.
Here, $k_{\rm F}$ is the ``average'' Fermi momentum of the
whole system. The kinetic density in momentum space for each
particle species is given by
\begin{equation}
\tilde\tau_{q \sigma} (k)
= k^2 \, \tilde\rho_{q \sigma} (k) ,
\end{equation}
and the kinetic density in coordinate space is
\begin{equation}
\tau_{q \sigma}
= \tfrac{V}{10 \pi} k_{\rm F,q \sigma}^5
= \tfrac{3}{20} \, \kinfac \, \rho_{q \sigma}^{5/3}
\quad ,
\end{equation}
where $\kinfac = ( 3 \pi^2 / 2 )^{2/3}$.
Various kinetic densities in the spin-isospin space are given by
\begin{eqnarray}
\tau_0
& = & \taunu + \taund + \taupu + \taupd
   =   \tfrac{3}{5} \, \kinfac \, \rho_0^{5/3}
       F_{5/3}^{(0)}
       \nn \\
\tau_1
& = & \taunu + \taund - \taupu - \taupd
   =   \tfrac{3}{5} \, \kinfac \, \rho_0^{5/3}
       F_{5/3}^{(\tau)}
       \nn \\
T_0
& = & \taunu - \taund + \taupu - \taupd
   =   \tfrac{3}{5} \, \kinfac \, \rho_0^{5/3}
       F_{5/3}^{(\sigma)}
       \nn \\
T_1
& = & \taunu - \taund - \taupu + \taupd
   =   \tfrac{3}{5} \, \kinfac \, \rho_0^{5/3}
       F_{5/3}^{(\sigma\tau)} ~,
\end{eqnarray}
where $F_m^{(0)}$, $F_m^{(\tau)}$, $F_m^{(\sigma)}$, and
$F_m^{(\sigma\tau)}$  are functions of the relative excesses:
\begin{eqnarray}
F_m^{(0)}
&  = & \tfrac{1}{4}
        \big[
          \big( 1 + I_\tau + I_\sigma + I_{\sigma \tau} \big)^m
        + \big( 1 + I_\tau - I_\sigma - I_{\sigma \tau} \big)^m
        \nn \\
&    & + \big( 1 - I_\tau + I_\sigma - I_{\sigma \tau} \big)^m
        + \big( 1 - I_\tau - I_\sigma + I_{\sigma \tau} \big)^m
        \big]
        \nn \\
\end{eqnarray}
This is a straightforward generalization of the corresponding definition
for asymmetric unpolarized nuclear matter given in \cite{Cha97a}.
Similarly one defines
\begin{eqnarray}
F_m^{(\tau)}
&  = & \tfrac{1}{4}
        \big[
          \big( 1 + I_\tau + I_\sigma + I_{\sigma \tau} \big)^m
        + \big( 1 + I_\tau - I_\sigma - I_{\sigma \tau} \big)^m
        \nn \\
&    & - \big( 1 - I_\tau + I_\sigma - I_{\sigma \tau} \big)^m
        - \big( 1 - I_\tau - I_\sigma + I_{\sigma \tau} \big)^m
        \big] ,
        \nn \\
F_m^{(\sigma)}
&  = & \tfrac{1}{4}
        \big[
          \big( 1 + I_\tau + I_\sigma + I_{\sigma \tau} \big)^m
        - \big( 1 + I_\tau - I_\sigma - I_{\sigma \tau} \big)^m
        \nn \\
&    & + \big( 1 - I_\tau + I_\sigma - I_{\sigma \tau} \big)^m
        - \big( 1 - I_\tau - I_\sigma + I_{\sigma \tau} \big)^m
        \big] ,
        \nn \\
F_m^{(\sigma\tau)}
&  = & \tfrac{1}{4}
        \big[
          \big( 1 + I_\tau + I_\sigma + I_{\sigma \tau} \big)^m
        - \big( 1 + I_\tau - I_\sigma - I_{\sigma \tau} \big)^m
        \nn \\
&    & - \big( 1 - I_\tau + I_\sigma - I_{\sigma \tau} \big)^m
        + \big( 1 - I_\tau - I_\sigma + I_{\sigma \tau} \big)^m
        \big] .
        \nn \\
\end{eqnarray}
For calculations of INM properties, we also need derivatives of
these functions. The first derivatives are given by
\begin{eqnarray}
\frac{\partial F_m^{(\tau)}}{\partial I_\tau}
& = & \frac{\partial F_m^{(\sigma)}}{\partial I_\sigma}
   =   \frac{\partial F_m^{(\sigma\tau)}}{\partial I_{\sigma \tau}}
   =   m \, F_{m-1}^{(0)} ,
       \nn \\
\frac{\partial F_m^{(0)}}{\partial I_\tau}
& = & \frac{\partial F_m^{(\sigma)}}{\partial I_{\sigma \tau}}
   =   \frac{\partial F_m^{(\sigma\tau)}}{\partial I_\sigma}
   =   m \, F_{m-1}^{(\tau)} ,
       \nn \\
\frac{\partial F_m^{(0)}}{\partial I_\sigma}
& = & \frac{\partial F_m^{(\tau)}}{\partial I_{\sigma \tau}}
   =   \frac{\partial F_m^{(\sigma\tau)}}{\partial I_\tau}
   =   m \, F_{m-1}^{(\sigma)} ,
       \nn \\
\frac{\partial F_m^{(0)}}{\partial I_{\sigma \tau}}
& = & \frac{\partial F_m^{(\tau)}}{\partial I_\sigma}
    =  \frac{\partial F_m^{(\sigma)}}{\partial I_\tau}
   =   m \, F_{m-1}^{(\sigma\tau)} ,
\end{eqnarray}
while the second derivatives are
\begin{eqnarray}
\frac{\partial^2 F_m^{(0)}}{\partial I_i^2}
& = & m (m-1) \, F_{m-2}^{(0)} ,
       \nn \\
\frac{\partial^2 F_m^{(j)}}{\partial I_i^2}
& = & m (m-1) \, F_{m-2}^{(j)} ,
\end{eqnarray}
for any \mbox{$i,j=\tau, \sigma, \sigma\tau$}.
Functions of the order \mbox{$m=0$} and \mbox{$m=1$} are rather simple:
\begin{eqnarray}
F_0^{(0)}
& = & 1
\quad , \quad
F_0^{(i)}
   =   0 \quad ,
       \nn \\
F_1^{(0)}
& = & 1
\quad , \quad
F_1^{(i)}
   =   I_i \quad ,
\end{eqnarray}
for any \mbox{$i=\tau, \sigma, \sigma\tau$}. Some special values
$F_m^{(i)}(I_\tau,I_\sigma,I_{\sigma\tau})$ appearing in limiting
cases of INM are
\begin{eqnarray}
F_m^{(0)}(0,0,0)
& = & 1
\quad , \quad
F_m^{(i)}(0,0,0)
   =  0
      \nn \\
F_m^{(0)}(1,0,0)
& = &
F_m^{(0)}(0,1,0)
   =
F_m^{(0)}(0,0,1)
   = 2^{m-1}
     \nn \\
F_m^{(\tau)}(1,0,0)
& = &
F_m^{(\sigma)}(0,1,0)
   =
F_m^{(\sigma\tau)}(0,0,1)
   =  2^{m-1}
     \nn \\
F_m^{(0)}(1,1,1)
& = &
F_m^{(1)}(1,1,1)
   =  4^{m-1} \quad .
\end{eqnarray}
while \mbox{$F_m^{(i)} = 0$} if \mbox{$I_i = 0$} and one of the
other $I_j$'s is equal to 1, with the last equal to zero.
These functions are useful  when writing down
the equation of state and its derivatives.
%
%
\subsection{``Equation of state'' of asymmetric polarized nuclear matter}
In INM
$\Delta \rho_{t t_3} (\rvec) 
= \Delta \vec{s}_{t t_3} (\rvec) 
(\rvec)
= \vec{j}_{t t_3} (\rvec) 
= \tensor{J}_{t t_3}  (\rvec) 
= 0$.
We choose pure neutron and proton states, which leads to
$\rho_{1, \pm 1} = 0$, $\rho_1 := \rho_{1,0}$, and similarly for all
other densities. We take the $z$ axis as the quantization axis for
the spin, i.e., $s_{t, x} = s_{t, y} = 0$, $s_t := s_{t, z}$, and for
the kinetic spin density $\vec{T}$. As discussed in Refs.\ \cite{Dab72a},
this breaks the isotropy of INM, leading to an axially deformed Fermi
surface, an effect which we neglect. Adding the kinetic term, the total
energy per nucleon (i.e.\ the ``equation of state'') for  the energy
functional (\ref{eq:SkyrmeFu:even}) and (\ref{eq:SkyrmeFu:odd}) is given by
\begin{eqnarray}
\label{eq:EOS}
\frac{{\cal H}}{\rho_0}
& = & \tfrac{3}{5}
       \tfrac{\hbar^2}{2m} \, \kinfac \, \rho_0^{2/3} \,
       F_{5/3}^{(0)}
       \nn \\
&   &       + \big(
             C_0^{\rho}
           + C_1^{\rho} \, I_\tau^2
           + C_0^{s} \, I_\sigma^2
           + C_1^{s} \, I_{\sigma \tau}^2
         \big) \, \rho_0
        \nn \\
&   &  + \tfrac{3}{5} \big(
             C_0^{\tau}                 F^{(0)}_{5/3}
           + C_1^{\tau} I_\tau          F^{(\tau)}_{5/3}
        \nn \\
&   &
           + C_0^{T}    I_\sigma        F^{(\sigma)}_{5/3}
           + C_1^{T}    I_{\sigma \tau} F^{(\sigma\tau)}_{5/3}
         \big) \,
       \kinfac \, \rho_0^{5/3} \, .
\end{eqnarray}
For unpolarized INM one has $I_\sigma = I_{\sigma \tau} = 0$ which
recovers the expression given in Ref.\cite{Cha97a}.

An interesting special case is polarized neutron matter, which
is discussed in \cite{Kut94a} for the Skyrme interactions. A stability
criterion derived there from the two-body force point of view
as outlined in Appendix \ref{Sect:app:SF} was used to constrain
the parameters of the
SLy$x$ forces \cite{Cha97a,Cha98a}. In this limiting case, one has
$\rhonu = \rho_0$, $\rhond = \rhopu = \rhopd = 0$, which is equivalent
to $I_\tau = I_\sigma = I_{\sigma \tau} = 1$ and leads to
\begin{eqnarray}
\label{eq:INM:PNM:Sfunc}
\frac{{\cal H}}{\rho_0}
& = & 2^{4/3} \kinfac \tfrac{3}{5}
       \left[ \tfrac{\hbar^2}{2m}
       + \big( C_0^{\tau} + C_1^{\tau} + C_0^{T} + C_1^{T} \big) \rho_0
       \right]
       \rho_0^{2/3}
       \nn \\
&   & \quad
       + \big( C_0^{\rho} + C_1^{\rho} + C_0^{s} + C_1^{s} \big) \,
         \rho_0 \quad .
\end{eqnarray}
Expressions (\ref{tx2BC}) for an antisymmetrized Skyrme force imply that
$C_0^{\rho} + C_1^{\rho} + C_0^{s} + C_1^{s} = 0$, and
\begin{equation}
\label{eq:INM:PNM:Sforce}
\frac{{\cal H}}{\rho_0}
   =   2^{4/3} \kinfac \tfrac{3}{5}
       \left[   \tfrac{\hbar^2}{2m}
              + \tfrac{1}{2} \, t_2 \, (1 + x_2) \, \rho_0
       \right]
       \rho_0^{2/3}
\quad .
\end{equation}
The stability of polarized neutron matter for all densities requires
\mbox{$x_2 \approx -1$} \cite{Kut94a}, so the SLy$x$ interactions take
\mbox{$x_2 \equiv - 1$}
\cite{Cha97a,Cha98a}. However, from the energy-density-functional
point of view, the coupling constants are independent, and the second
term in Eq.\ (\ref{eq:INM:PNM:Sfunc}) also contributes to the
stability condition.
%
%
\subsection{Pressure, Incompressibility and Asymmetry Coefficients}
At the saturation point, all first derivatives of the energy per nucleon
have to vanish and all second derivatives have to be positive. The first
derivative with respect to $\rho_0$ is related to the pressure, the second
derivative with respect to $\rho_0$ is related to the incompressibility,
and the second  derivatives with respect to the $I_i$ is related to the
asymmetry coefficients.  For symmetric matter, the first derivatives with
respect to the $I_i$ vanish  because the energy per nucleon is an even
function of all $I_i$s.
The pressure is given by
\begin{equation}
P
= - \frac{\partial E}{\partial V} \bigg|_A
= \rho_0^2 \frac{\partial \HF/\rho_0}{\partial \rho_0}
\quad ,
\end{equation}
which gives
\begin{eqnarray}
P
& = & \tfrac{2}{5}
       \tfrac{\hbar^2}{2m}
       \, \kinfac \, \rho_0^{5/3}
       F_{5/3}^{(0)}
       \nn \\
&   & \quad
       + \big(   C_0^{\rho}
               + C_1^{\rho} I_\tau^2
               + C_0^{s} I_\sigma^2
               + C_1^{s} I_{\sigma \tau}^2
         \big) \, \rho_0^2
       \nn \\
&   & \quad
       + 3 \rho_0^2 \; \frac{\partial}{\partial \rho_0} \,
         \big(   C_0^{\rho}
               + C_1^{\rho} I_\tau^2
               + C_0^{s} I_\sigma^2
               + C_1^{s} I_{\sigma \tau}^2
         \big)
       \nn \\
&   & \quad
       + \kinfac
         \Big(   C_0^{\tau} \, F^{(0)}_{5/3}
               + C_1^{\tau} \, F^{(\tau)}_{5/3} \, I_\tau
       \nn \\
&   & \qquad
               + C_0^{T} \, F^{(\sigma)}_{5/3} \,  I_\sigma
               + C_1^{T} \, F^{(\sigma\tau)}_{5/3} \,  I_{\sigma \tau}
         \Big) \; \rho^{8/3}
\quad .
\end{eqnarray}
The incompressibility is defined as
\begin{equation}
K
= \frac{18 P}{\rho_0}
   + 9 \rho_0^2 \frac{\partial^2 \HF/\rho_0}{\partial \rho_0^2} ,
\end{equation}
which, for the Skyrme energy functional (\ref{eq:EOS}) at the
saturation point (\mbox{$\rho_0=\rho_{\rm n.m.}$},
$I_\tau = I_\sigma = I_{\sigma \tau} = 0$) gives
\begin{eqnarray}
K_\infty
& = & - \tfrac{6}{5}
       \left(   \tfrac{\hbar^2}{2m}
              - 5 C_0^\tau \, \rho_0
       \right)
       \kinfac \, \rho_0^{2/3}
       F_{5/3}^{(0)}
       \nn \\
&   & + 2 \rho_0^2 \frac{\partial C_0^\rho}{\partial \rho_0}
       + \rho_0^3   \frac{\partial^2 C_0^\rho}{\partial^2 \rho_0} .
\end{eqnarray}
The asymmetry coefficients are:
\begin{eqnarray}
a_\tau
& = & \frac{1}{2}
       \frac{\partial^2 \HF/\rho_0}{\partial I_\tau^2}
       \bigg|_{I_\tau = I_\sigma = I_{\sigma \tau} = 0}
       \nn \\
& = & \tfrac{1}{3}
       \left[   \tfrac{\hbar^2}{2m}
              + ( C_0^{\tau} + 3 C_1^{\tau} ) \, \rho_0
       \right]
       \kinfac \, \rho_0^{2/3}
       + C_1^{\rho} \, \rho_0 ,
      \\
a_\sigma
& = & \frac{1}{2}
       \frac{\partial^2 \HF/\rho_0}{\partial I_\sigma^2}
       \bigg|_{I_\tau = I_\sigma = I_{\sigma \tau} = 0}
       \nn \\
& = & \tfrac{1}{3}
       \left[   \tfrac{\hbar^2}{2m}
              + ( C_0^{\tau} + 3 C_0^T) \, \rho_0
       \right]
       \kinfac \, \rho_0^{2/3}
       + C_0^{s} \, \rho_0 ,
      \\
a_{\sigma \tau}
& = & \frac{1}{2}
       \frac{\partial^2 \HF/\rho_0}{\partial I_{\sigma \tau}^2}
       \bigg|_{I_\tau = I_\sigma = I_{\sigma \tau} = 0}
       \nn \\
& = & \tfrac{1}{3}
       \left[ \tfrac{\hbar^2}{2m}
              + ( C_0^{\tau} + 3 C_1^T ) \, \rho_0 \right]
       \kinfac \, \rho_0^{2/3}
       + C_1^{s} \, \rho_0  .
\end{eqnarray}
Here, $a_\tau$ is the well-known volume asymmetry coefficient of the
liquid-drop model, and $a_\sigma$ and $a_{\sigma \tau}$ are its
generalizations to the spin and spin-isospin channels of the interaction.
At the saturation point, all asymmetry coefficients have to be positive.
%
%
\section{Landau Parameters from the Skyrme energy functional}
\label{Sect:app:landau}
A simple and instructive description of the residual interaction
in homogeneous INM is given by the Landau interaction developed in the
context of Fermi-liquid theory \cite{Tow87a}. Landau parameters
corresponding to the Skyrme forces are discussed in Refs.\
\cite{Fri86a,Liu91a,Kre77a,War83a,Gia81a,Bae75a}.
Starting from the full density matrix in (relative) momentum space
$\tilde\rho (\kvec \sigma \tau \sigma' \tau')$, the various densities
are defined as
\begin{mathletters}
\begin{eqnarray}
\tilde\rho_{00} (\kvec)
& = & \sum_{\sigma} \sum_{\tau}
       \tilde\rho (\kvec \sigma \tau \sigma \tau ) ,
       \\
\tilde\rho_{1 t_3} (\kvec)
& = & \sum_{\sigma} \sum_{\tau,\tau'}
       \tilde\rho (\kvec \sigma \tau \sigma \tau' ) \;
       \tau^{t_3}_{\tau \tau'} ,
       \\
\tilde{\vec{s}}_{00} (\kvec)
& = & \sum_{\sigma, \sigma'} \sum_{\tau}
       \tilde\rho (\kvec \sigma \tau \sigma' \tau ) \;
       \sigmavec_{\sigma \sigma'} ,
       \\
\tilde{\vec{s}}_{1 t_3} (\kvec)
& = & \sum_{\sigma, \sigma'} \sum_{\tau, \tau'}
       \tilde\rho (\kvec \sigma \tau \sigma' \tau' ) \:
       \sigmavec_{\sigma \sigma'} \; \tau^{t_3}_{\tau \tau'} ,
\end{eqnarray}
\end{mathletters}

The kinetic densities are given by \mbox{$\tau_{t t_3} = \rho_{t t_3} \,
k^2$},
\mbox{$\vec{T}_{t t_3} = \vec{s}_{t t_3} \, k^2$}. The Landau-Migdal
interaction is defined as
\begin{eqnarray}
\lefteqn{
\tilde{F}
(\kvec_1 \sigma_1 \tau_1 \sigma'_1 \tau'_1;
   \kvec_2 \sigma_2 \tau_2 \sigma'_2 \tau'_2 )
} \nn \\
& = & \frac{\delta^2 \EF}
            {\delta \tilde\rho (\kvec_1 \sigma_1 \tau_1 \sigma'_1 \tau'_1)
             \delta \tilde\rho (\kvec_2 \sigma_2 \tau_2 \sigma'_2 \tau'_2)}
       \nn \\
& = & \tilde{f} (\kvec_1, \kvec_2)
       + \tilde{f}{}' (\kvec_1, \kvec_2) \;
         \bbox{\tau}_1 \cdot \bbox{\tau}_2
         \phantom{\frac{\delta^2}{\tilde\rho}}
       \nn \\
&   & +
         \tilde{g} (\kvec_1, \kvec_2) \; \bbox{\sigma}_1 \cdot \bbox{\sigma}_2
       + \tilde{g}{}' (\kvec_1, \kvec_2) \,
        (\bbox{\sigma}_1 \cdot \bbox{\sigma}_2)
         (\bbox{\tau}_1 \cdot \bbox{\tau}_2) .
         \phantom{\bigg|}
\end{eqnarray}
The isoscalar-scalar, isovector-scalar, isoscalar-vector, and
isovector-vector channels of the residual interaction are given by
\begin{mathletters}
\begin{eqnarray}
\tilde{f} (\kvec_1, \kvec_2)
& = & \frac{\delta^2 \EF}
            {\delta \tilde\rho_{00} (\kvec_1) \delta \tilde\rho_{00}
(\kvec_2)}
       \\
\tilde{f}{}' (\kvec_1, \kvec_2)
& = & \frac{\delta^2 \EF}
            {\delta \tilde\rho_{1 t_3} (\kvec_1)
             \delta \tilde\rho_{1 t_3} (\kvec_2)}
       \\
\tilde{g} (\kvec_1, \kvec_2)
& = &  \frac{\delta^2 \EF}
            {\delta \tilde{\vec{s}}_{00} (\kvec_1)
             \delta \tilde{\vec{s}}_{00} (\kvec_2)}
       \\
\tilde{g}{}' (\kvec_1, \kvec_2)
& = & \frac{\delta^2 \EF}
            {\delta \tilde{\vec{s}}_{1 t_3} (\kvec_1)
             \delta \tilde{\vec{s}}_{1 t_3} (\kvec_2)}
\end{eqnarray}
\end{mathletters}
Assuming that only states at the Fermi surface contribute, i.e.,
$| \kvec_1 | = | \kvec_2 | = k_{\rm F}$,
$\tilde{f}$, $\tilde{f}{}'$, $\tilde{g}$, and $\tilde{g}{}'$ depend on
the angle $\theta$ between
$\kvec_1$ and $\kvec_2$ only, and can be expanded into Legendre
polynomials, e.g.
\begin{equation}
\tilde{f} (\kvec_1, \kvec_2)
= \frac{1}{N_0} \sum_{\ell = 0}^{\infty} f_\ell \; P_\ell (\theta) .
\end{equation}
The normalization factor $N_0$ is the level density at the
Fermi surface
\begin{equation}
\label{eq:N0}
\frac{1}{N_0}
= \frac{\pi^2 \hbar^2}{2 m^* k_{\rm F} }
\approx 150 \, \frac{m}{m^*} \; \text{MeV} \; \text{fm}^3  
\quad .
\end{equation}
A variety of definitions of the normalization factor $N_0$ are used
in the literature and great care has to be taken when comparing values
from different groups; see, e.g.,\ Ref.\cite{Tow87a} for a detailed
discussion. We use the convention defined in \cite{Gia81a}.
The Landau parameters corresponding to the general energy
functional (\ref{eq:SkyrmeFu}) are
\begin{eqnarray}
f_0
& = &  N_0 \left( 2 C_0^{\rho}
                 + 4 \frac{\partial C_0^{\rho}}{\partial \rho_{00}} \rho_0
                 + \frac{\partial^2 C_0^{\rho}}{\partial \rho_{00}^2} \rho_0^2
                 + 2 C_0^{\tau} \, \kinfac \, \rho_{00}^{2/3}
             \right) ,
       \nn \\
f_0'
& = & N_0 \big(   2 C_1^{\rho}
                 + 2 C_1^{\tau} \, \kinfac \, \rho_{00}^{2/3}
           \big) ,
       \nn \\
g_0
& = & N_0 \big(   2 C_0^{s}
                 + 2 C_0^{T} \, \kinfac \, \rho_{00}^{2/3}
           \big) ,
       \nn \\
g_0
& = & N_0 \big(   2 C_1^{s}
                 + 2 C_1^{T} \, \kinfac \, \rho_{00}^{2/3}
           \big) ,
       \nn \\
f_1
& = & - 2 N_0 \; C_0^{\tau} \, \kinfac \, \rho_{00}^{2/3} ,
       \nn \\
f_1'
& = & - 2 N_0 \; C_1^{\tau} \, \kinfac \, \rho_0^{2/3} ,
       \nn \\
g_1
& = & - 2 N_0 \; C_0^{T} \, \kinfac \, \rho_{00}^{2/3},
       \nn \\
g_1'
& = & - 2 N_0 \; C_1^{T} \, \kinfac \, \rho_{00}^{2/3}.
\end{eqnarray}
Higher-order Landau parameters vanish for the second-order
energy functional (\ref{eq:SkyrmeFu:gauge}), but not for
finite-range interactions as the Gogny force discussed in
the next Appendix.
The Landau parameters provide a stability criterion for symmetric
unpolarized INM:  It becomes unstable for a given interaction
when either $f_\ell$, $f_\ell'$, $g_\ell$, or $g_\ell'$ is less
than $-(2 \ell + 1)$.
%
%
\section{Landau Parameters from the Gogny force}
\label{Sect:app:landau:gogny}
The residual interaction in INM from the Gogny force \cite{Dec80a}
\begin{eqnarray}
\label{Gognyforce}
\lefteqn{V_{\rm Gogny}(\vec{r}_1, \vec{r}_2)}
\nn \\
& = & \sum_{i=1,2} \big(
                       W_i
                     + B_i \hat{P}_\sigma
                     + H_i \hat{P}_\tau
                     - M_i \hat{P}_\sigma \hat{P}_\tau
                    \big) {\rm e}^{-(\vec{r}_{1}-\vec{r}_2)^2/\mu_i^2}
        \nn \\
&   & + t_0 ( 1 + x_0 \hat{P}_\sigma) \;
         \rho_0^\alpha \left(\tfrac{\vec{r}_1+\vec{r}_2}{2}\right) \,
         \delta (\vec{r}_1 - \vec{r}_2)
         \phantom{\sum_{i=1,2}}
       \nn \\
&   &
       + \iunit W_0 \,
         ( \hat{\sigmavec}_1 + \hat{\sigmavec}_2 ) \cdot
         \hat{\vec{k}}{}' \times
         \delta (\rvec_1 - \rvec_2) \,
         \hat{\vec{k}}
\end{eqnarray}
(see Appendix \ref{Sect:app:SF} for the definition of $\hat{\vec{k}}$,
$\hat{\vec{k}}{}^\prime$, $\hat{P}_\sigma$, and $\hat{P}_\tau$) has been
discussed in \cite{Gog77a,Ven94a}. Evaluating the expressions given in
\cite{Ven94a} for $(k, k', q) = (k_{\rm F}, k_{\rm F}, 0)$,  one obtains 
the usual Landau parameters
\begin{eqnarray}
f_\ell
& = & \sum_{i=1,2}
       \big[   ( 4 W_i + 2 B_i - 2 H_i -   M_i ) \Psi^{(i)}_\ell
       \nn \\
&   & \phantom{\sum}
              + ( - W_i - 2 B_i + 2 H_i + 4 M_i ) \Phi^{(i)}_\ell
       \big]
       \nn \\
&   & \phantom{\sum}
       + \delta_{\ell 0} \,
         \tfrac{3}{8} t_0 \, (\alpha + 1) (\alpha + 2) \, \rho_0^\alpha
       \nn \\
f_\ell'
& = & \sum_{i=1,2}
       \big[ - (                 2 H_i +   M_i )  \Phi^{(i)}_\ell
             - (   W_i - 2 B_i                 )  \Psi^{(i)}_\ell
       \big]
       \nn \\
&   & + \delta_{\ell 0} \,
         \tfrac{1}{4} t_0 \, (1 + 2 x_0) \rho_0^\alpha
       \nn \\
g_\ell
& = & \sum_{i=1,2}
       \big[   (         2 B_i         -   M_i )  \Psi^{(i)}_\ell
             + ( - W_i         + 2 H_i         )  \Phi^{(i)}_\ell
       \big]
       \nn \\
&   & + \delta_{\ell 0} \,
         \tfrac{1}{4} t_0 \, (1 - 2 x_0) \rho_0^\alpha
       \nn \\
g_\ell'
& = & - \sum_{i=1,2}
       \big( M_i \Psi^{(i)}_\ell + W_i \Phi^{(i)}_\ell \big)
       + \delta_{\ell 0} \, \tfrac{1}{4} t_0 \, \rho_0^\alpha
\end{eqnarray}
where
\begin{eqnarray}
\Psi^{(i)}_\ell
& = & \tfrac{1}{4} \pi^{3/2} \mu_i^3 \, N_0 \; \delta_{\ell 0}
       \nn \\
\Phi^{(i)}_0
& = & \tfrac{1}{4} \pi^{3/2} \mu_i^3 \, N_0 \;
       {\rm e}^{-z} \; \frac{\sinh (z)}{z}
       \nn \\
\Phi^{(i)}_1
& = & \tfrac{3}{4} \pi^{3/2} \mu_i^3 \, N_0 \; {\rm e}^{-z}
       \left(\frac{\cosh (z)}{z} - \frac{\sinh (z)}{z^2}\right)
       \nn \\
\Phi^{(i)}_2
& = & \tfrac{5}{4} \pi^{3/2} \mu_i^3 \, N_0 \; {\rm e}^{-z}
       \left[   \sinh(z) \left(\frac{1}{z} + \frac{3}{z^3} \right)
              - \frac{3 \cosh(z)}{z^2}
       \right]
      \nn
\end{eqnarray}
with \mbox{$z= \mu_i^2 k_{\rm F}^2 /2$}. The normalization factor
$N_0$ is again given by (\ref{eq:N0}).
%
%
\section{Residual Interaction in Finite Nuclei}
\label{Sect:app:resint}
Equation\ (\ref{eq:d2Edrhodrho}) gives the most general form of the
residual interaction in finite nuclei. Only a few terms contribute to
the $1^+$ isovector excitations of the even-even nuclei we are interested 
in. First of all, only the isovector densities contribute. Next, the
conditions \mbox{$\Delta J=1$} and \mbox{$\Delta \pi = 0$} between
ground state and excited states imply that the only terms in the
energy functional that can contribute are quadratic in local tensor
or vector parity-even densities/currents. As can be seen from
Table~2 in \cite{Dob96d}, all possible contributions are time-odd.
One finally obtains
\begin{eqnarray}
\label{eq:d2Edrhodrho2}
\lefteqn{
v_{\rm res} (\vec{r}_1, \vec{r}_2)
} \nn \\
& = & \frac{\delta^2 {\cal E}}
            {\delta \vec{s}_{1 t} (\vec{r}_1) \,
             \delta \vec{s}_{1 t} (\vec{r}_2)}
       \, (\bbox{\sigma}_1 \cdot \bbox{\sigma}_2) \,
          (\bbox{\tau}_1   \cdot \bbox{\tau}_2)
       \nn \\
& = & \Big[  2 C_1^{s} [\rho_{00}] \;  \delta (\rvec_1 - \rvec_2)
       \nn \\
&   & + \tfrac{1}{2} ( C_1^{T} - 4 C_1^{\Delta s} ) \;
        (   \hat{\vec{k}}{}^{\prime 2} \delta (\rvec_1 - \rvec_2)
          + \delta (\rvec_1 - \rvec_2)\hat{\vec{k}}{}^{2} )
       \nn \\
&   & + ( 3 C_1^{T} + 4 C_1^{\Delta s} ) \;
       \hat{\vec{k}}{}' \cdot \delta (\rvec_1 - \rvec_2) \, \hat{\vec{k}}
       \Big] \;
       \hat{\bbox{\sigma}}_1 \cdot  \hat{\bbox{\sigma}}_2 \;
       \hat{\bbox{\tau}}_1   \cdot  \hat{\bbox{\tau}}_2
       \nn \\
&   & - 2 \iunit C_1^{\nabla J} \;
       \hat{\bbox{\tau}}_1 \cdot \hat{\bbox{\tau}}_2 \;
       ( \hat{\bbox{\sigma}}_1 + \hat{\bbox{\sigma}}_2 ) \cdot
       \hat{\vec{k}}{}^\prime \times \delta (\rvec_1 - \rvec_2) \,
       \hat{\vec{k}}
       \nn
\end{eqnarray}
where $\hat{\vec{k}}$ and $\hat{\vec{k}}{}^\prime$ are defined in 
Appendix \ref{Sect:app:SF}. Since the coupling constants depend only
on the scalar isoscalar density $\rho_{00}$, there are no rearrangement
terms in the spin-isospin channel of the residual interaction.
Unsymmetrized proton-neutron matrix elements of this interaction are
to be inserted into the QRPA equations as outlined in Ref.~\cite{Eng99a}.
\end{appendix}
%
%


\begin{references}

\bibitem{Ring}
   P. Ring and  P. Schuck,
   \emph{The Nuclear Many-Body Problem}
   (Springer-Verlag, Berlin, 1980).

\bibitem{Eng75a}
   Y. M. Engel, D. M. Brink, K. Goeke, S. J. Krieger, and D. Vautherin,
   Nucl. Phys. \textbf{A249},  215  (1975).

\bibitem{Dob95c}
   J. Dobaczewski and J. Dudek,
   Phys. Rev. C \textbf{52}, 1827    (1995),
   Phys. Rev. C \textbf{55}, 3177(E) (1997).

\bibitem{Sat98}
   W. Satu{\l}a,
   Proc. Nuclear Structure '98,
   Gatlinburg, Tennessee, August 1998,
   C. Baktash [ed.],
   AIP Conference Proceedings \textbf{481}, 141 (1999).

\bibitem{Xu99a}
   R. R. Xu, R. Wyss, and P. M. Walker,
   Phys. Rev. C \textbf{60},  051301  (1999).

\bibitem{Rut99a}
   K. Rutz, M. Bender, P.--G. Reinhard, and J. A. Maruhn,
   Phys. Lett. \textbf{B468},  1 (1999).

\bibitem{Ber80a}
   V. Bernard and Nguyen Van Giai
   Nucl. Phys. \textbf{A348}, 75 (1980).

\bibitem{Rut98a}
   K. Rutz, M. Bender, P.--G. Reinhard, J. A. Maruhn, and W. Greiner,
   Nucl. Phys. \textbf{A634},  67  (1998).

\bibitem{Mol00}
   H. Molique, J. Dobaczewski, and J. Dudek,
   Phys. Rev. C \textbf{61}, 044304 (2000).

\bibitem{Lip77a}
   E. Lipparini, S. Stringari, and M. Traini,
   Nucl. Phys. \textbf{A293}, 29 (1977).

\bibitem{Eng99a}
   J. Engel, M. Bender, J. Dobaczewski, W. Nazarewicz, and R. Surman,
   Phys. Rev. C \textbf{60}, 014302 (1999).

\bibitem{Rin96}
   P. Ring,
   Prog. Part. Nucl. Phys. \textbf{37}, 193 (1996).

\bibitem{Rei89a}
   P.--G. Reinhard,
   Rep. Prog. Phys. \textbf{52}, 439 (1989).

\bibitem{Afa00}
   A. V. Afanasjev and P. Ring,
   Phys. Rev. C \textbf{62}, 031302 (2000).

\bibitem{Hoh64a}
   P. Hohenberg, W. Kohn,
   Phys. Rev. \textbf{B136},  864  (1964);\\
   W. Kohn, L. J. Sham,
   Phys. Rev. \textbf{A140},  1133  (1965);
   W. Kohn,
   Rev. Mod. Phys. \textbf{71}, 1253 (1998).

\bibitem{Nag98a}
   R. M. Dreizler, E. K. U. Gross,
   \emph{Density functional theory},
   Springer, Berlin 1990;\\
   {\'A}. Nagy,
   Phys. Rep. \textbf{311}, 47 (1999);\\
   Ranbir Singh, B. M. Deb,
   Phys. Rep. \textbf{298}, 1 (1998).

\bibitem{Pet91aB}
   I. Z. Petkov and M. V. Stoitsov,
   \emph{Nuclear Density Functional Theory}
   (Clarendon Press, Oxford, 1991).

\bibitem{Neg72a}
   J. W. Negele and D. Vautherin,
   Phys. Rev. C \textbf{5},  1472 (1972);
   Phys. Rev. C \textbf{11}, 1031 (1975).

\bibitem{Dob84a}
   J. Dobaczewski, H. Flocard, and J. Treiner,
   Nucl. Phys. \textbf{A422}, 103 (1984).

\bibitem{Dob96d}
   J. Dobaczewski and J. Dudek,
   in \emph{High Angular Momentum Phenomena, Workshop in honour of
   Zdzis{\l}aw Szyma{\'n}ski, Piaski, Poland, August 23--26, 1995},
   Acta Phys. Pol. \textbf{B27}, 45 (1996).

\bibitem{Fri86a}
   J. Friedrich and P.--G. Reinhard,
   Phys. Rev. C \textbf{33},  335  (1986).

\bibitem{Bla95a}
   J. P. Blaizot, J. F. Berger, J. Decharg{\'e}, and M. Girod,
   Nucl. Phys. \textbf{A591}, 435 (1995).

\bibitem{Cha97a}
   E. Chabanat, P. Bonche, P. Haensel, J. Meyer, and R. Schaeffer,
   Nucl. Phys. \textbf{A627}, 710 (1997).

\bibitem{Cha98a}
   E. Chabanat, P. Bonche, P. Haensel, J. Meyer, and R. Schaeffer,
   Nucl. Phys. \textbf{A635}, 231 (1998),
   Nucl. Phys. \textbf{A643}, 441(E) (1998).

\bibitem{Sky56a}
   T. H. R. Skyrme,
   Philos. Mag. \textbf{1}, 1043 (1956);
   Nucl. Phys. \textbf{9}, 615 (1959).

\bibitem{Sta77}
   Fl. Stancu, D.M. Brink, and H. Flocard,
   Phys. Lett. \textbf{68B}, 108 (1977).

\bibitem{Liu91a}
   K. Liu, H. Luo, Z. Ma, Q. Shen, and S. A. Moszkowski,
   Nucl. Phys. \textbf{A534}, 1  (1991);\\
   K. Liu, H. Luo, Z. Ma, and Q. Shen,
   \emph{ibid}, 25;\\
   K. Liu, H. Luo, Z. Ma, M. Feng, and Q. Shen,
   \emph{ibid}, 48;\\
   K. Liu, Z. Ma, and H. Luo,
   \emph{ibid}, 58.

\bibitem{Ton84a}
   F. Tondeur, M. Brack, M. Farine, and J. M. Pearson,
   Nucl. Phys. \textbf{A420}, 297 (1984).

\bibitem{Bro98a}
   B. A. Brown,
   Phys. Rev. C \textbf{58}, 220 (1998).

\bibitem{Bei75a}
   M. Beiner, H. Flocard, Nguyen Van Giai, and P. Quentin,
   Nucl. Phys. \textbf{A238}, 29 (1975).

\bibitem{PGRpc}
   P.--G. Reinhard,
   private communication.

\bibitem{Ton83a}
   F. Tondeur,
   Phys. Lett. \textbf{123B}, 139 (1983).

\bibitem{Vau72a}
   D. Vautherin and D. M. Brink,
   Phys. Rev. C \textbf{5}, 626  (1972).

\bibitem{Cha75a}
   B. D. Chang,
   Phys. Lett. \textbf{56B}, 205 (1975).

\bibitem{Str76a}
   S. Stringari, R. Leonardi, and D. M. Brink,
   Nucl. Phys. \textbf{A269}, 87 (1976).

\bibitem{Kre77a}
   S. Krewald, V. Klemt, J. Speth, and A. Faessler,
   Nucl. Phys. \textbf{A281}, 166 (1977).

\bibitem{War83a}
   M. Waroquier, K. Heyde, and G. Wenes,
   Nucl. Phys. \textbf{A404}, 269 (1983);
   M. Waroquier, G. Wenes, and K. Heyde,
   Nucl. Phys. \textbf{A404}, 298 (1983).

\bibitem{Gia81a}
   Nguyen Van Giai and H. Sagawa,
   Phys. Lett. \textbf{106B}, 379 (1981).

\bibitem{Rei85a}
   P.--G. Reinhard and J. Friedrich,
   Z. Phys. \textbf{A321}, 619 (1985);
   P.--G. Reinhard, M. Brack, and O. Genzgen,
   Phys. Rev. \textbf{A41},  5568  (1990);
   P.--G. Reinhard,
   Ann. Phys. (Leipzig) \textbf{1},  632  (1992).

\bibitem{Lal94a}
   G. A. Lalazissis, M. M. Sharma, J. K{\"o}nig, and P. Ring,
   Proc. of the Int. Conf. on
   ``Nuclear Shapes and Nuclear Structure at Low Excitation Energies'',
   Antibes (France) June 20--25, 1994,
   M. Vergnes, D. Goutte, P.--H. Heenen, and J. Sauvage [eds.]
   (Editions Frontieres, Gif-sur-Yvette Cedex, France, 1994), p. 161.

\bibitem{Sha95a}
   M. M. Sharma, G. A. Lalazissis, J. K{\"o}nig, and P. Ring,
   Phys. Rev. Lett. \textbf{74}, 3744 (1995).

\bibitem{Rei95a}
   P.--G. Reinhard and H. Flocard,
   Nucl. Phys. \textbf{A584}, 467 (1995).

\bibitem{Bel56a}
   J. S. Bell and T. H. R. Skyrme,
   Philos. Mag. \textbf{1}, 1055 (1956).

\bibitem{Nak82a}
   K. Nakayama , A. Pio Gale{\~a}o, and F. Krmpoti{\'c},
   Phys. Lett. \textbf{114B}, 217 (1982).

\bibitem{Ber81a}
   G. Bertsch, D. Cha, and H. Toki,
   Phys. Rev. C \textbf{24}, 533 (1981).

\bibitem{Gaa81aE}
   C. Gaarde, J. Rapaport, T. N. Taddeucci, C. D. Goodman, C. C. Foster,
   D. E. Bainum, C. A. Goulding, M. B. Greenfield, D. J. Horen,
   and E. Sugarbaker,
   Nucl. Phys. \textbf{A369}, 258 (1981).

\bibitem{Suz82a}
   T. Suzuki,
   Nucl. Phys. \textbf{A379}, 110 (1982).

\bibitem{Ber81b}
   G. F. Bertsch,
   Nucl. Phys. \textbf{A354}, 157c (1981).

\bibitem{Eng88a}
   J. Engel, P. Vogel, M. R. Zirnbauer,
   Phys. Rev. C \textbf{37}, 731 (1988).

\bibitem{Tow87a}
   I. S. Towner,
   Phys. Rep. \textbf{155},  263  (1987).

\bibitem{Ost91a}
   F. Osterfeld,
   \emph{Giant Gamow-Teller resonances}
   in ``Electric and magnetic giant resonances in nuclei''.
   J. Speth [ed.],
   International Review of Nuclear Physics Vol. 7,
   World Scientific (1991), page 536.

\bibitem{Ost92a}
   F. Osterfeld,
   Rev. Mod. Phys. \textbf{64}, 491 (1992).

\bibitem{Borzov}
  I. N. Borzov, S. A. Fayans, E. L. Trykov,
  Nucl. Phys. \textbf{A584}, 335 (1995);
  I. N. Borzov, S. A. Fayans, E. Kromer, D. Zawisha,
  Z. Phys. \textbf{A335}, 127 (1996);
  I. N. Borzov, S. Goriely, J. M. Pearson,
  Nucl. Phys. \textbf{A621}, 307c (1997);
  I. N. Borzov, S. Goriely, 
  Phys. Rev. C \textbf{62}, 035501 (2001).

\bibitem{Mol90a}
   P. M{\"o}ller and J. Randrup,
   Nucl. Phys. \textbf{A514}, 1 (1990).

\bibitem{Homma}
   H. Homma, E. Bender, M. Hirsch, K. Muto, H. V. Klapdor--Kleingrothaus,
   and T. Oda,
   Phys. Rev. C \textbf{54}, 2972 (1996); \\
   M. Hirsch, A. Staudt, and H. V. Klapdor--Kleingrothaus,
   At. Data Nucl. Data Tables \textbf{44}, 79 (1992); \\
   M. Hirsch, A. Staudt, K. Muto, and H. V. Klapdor--Kleingrothaus,
   At. Data Nucl. Data Tables \textbf{53}, 165 (1993).

\bibitem{Sar98a}
   P. Sarriguren, E. M. de Guerra, A. Escuderos, and A. C. Carrizo,
   Nucl. Phys. \textbf{A635}, 55 (1998);
   P. Sarriguren, E. M. de Guerra, and A. Escuderos,
   Nucl. Phys. \textbf{A658},  13 (1999);
   Nucl. Phys. \textbf{A691}, 631 (2001).

\bibitem{Sar96a}
   P. Sarriguren, E. M. de Guerra, and R. Nojarov,
   Phys. Rev. C \textbf{54}, 690 (1996);
   Z. Phys. \textbf{A357}, 143 (1997).

\bibitem{SkO}
   P.--G. Reinhard, D. J. Dean, W. Nazarewicz, J. Dobaczewski,
   J. A. Maruhn, and M. R. Strayer,
   Phys. Rev. C \textbf{60}, 014316 (1999).

\bibitem{Dec80a}
   J. Decharg{\'e} and D. Gogny,
   Phys. Rev. C \textbf{21}, 1568 (1980).

\bibitem{Wak97aE}
   T. Wakasa, H. Sakai, H. Okamura, H. Otsu, S. Fujita, S. Ishida,
   N. Sakamoto, T. Uesaka, Y. Satou, M. B. Greenfield, and K. Hatanaka,
   Phys. Rev. C \textbf{55}, 2909 (1997).

\bibitem{Ben99a}
   M. Bender, K. Rutz, P.--G. Reinhard, J. A. Maruhn, and W. Greiner,
   Phys. Rev. C \textbf{60}, 034304  (1999).

\bibitem{Sag01a}
   H. Sagawa and S. Yoshida,
   Nucl. Phys. \textbf{A688}, 755 (2001).

\bibitem{Spe77a}
   J. Speth, E. Werner, and W. Wild,
   Phys. Rep. \textbf{33}, 127  (1977).

\bibitem{Bro90a}
   R. Brockmann and R. Machleidt,
   Phys. Rev. C \textbf{42}, 1965 (1990).

\bibitem{Abe90a}
  S. \AA berg, H. Flocard, and W. Nazarewicz,
   Ann. Rev. Nucl. Part. Sci. \textbf{40}, 439 (1990).

\bibitem{Jan91}
   R. V. F. Janssens and T. L. Khoo,
   Ann. Rev. Nucl. Part. Sci. \textbf{41}, 321 (1991).

\bibitem{Bak95}
  C. Baktash, B. Haas, and W. Nazarewicz,
   Ann. Rev. Nucl. Part. Sci. \textbf{45}, 485 (1995).

\bibitem{Dob00d}
  J. Dobaczewski and J. Dudek,
   Comput. Phys. Commun. \textbf{102}, 166 (1997);
   \textbf{102}, 183 (1997);
   \textbf{131}, 164 (2000).

\bibitem{Hae89a}
   P. Haensel, J. L. Zdunik, and J. Dobaczewski,
   Astron. Astrophys. \textbf{222}, 353 (1989).

\bibitem{Kut94a}
   M. Kutschera and W. W{\'o}jcik,
   Phys. Lett. \textbf{B325}, 172 (1994).

\bibitem{Hae82a}
   P. Haensel and A. J. Jerzak,
   Phys. Lett. \textbf{112B}, 285 (1982).

\bibitem{Dab72a}
   J. D\c{a}browski and P. Haensel,
   Phys. Lett. \textbf{42B}, 163 (1972);
   Ann. Phys. (NY) \textbf{97}, 452 (1976); 
   P. Haensel and J. D\c{a}browski,
   Z. Phys. \textbf{A274}, 377 (1975); 
   Nucl. Phys. \textbf{A254}, 211 (1975).

\bibitem{Bae75a}
   S. O. B{\"a}ckman, A. D. Jackson, and J. Speth,
   Phys. Lett. \textbf{56B},  209 (1975).

\bibitem{Gog77a}
   D. Gogny and R. Padjen,
   Nucl. Phys. \textbf{A293} (1977) 365.\\
   Note that the D1 values for $M_2$, $H_2$, and $B_2$ in Table 8
   are interchanged and that the values for \mbox{$g_0' = F_0^{11}$}
   and \mbox{$g_2' = F_2^{11}$} in Table 1 contain typos.

\bibitem{Ven94a}
   J. Ventura, A. Polls, X. Vi{\~n}as, E. S. Hernandez,
   Nucl. Phys. \textbf{A578} (1994) 147.

\end{references}
\end{document}